\documentclass[12pt,letterpaper]{article}
\usepackage[pdftex]{graphicx,color}
\pdfoutput=1
\input{epsf}
\usepackage{amsmath,amssymb}
\usepackage[dvips,letterpaper,text={6.5in,9in}]{geometry}
\usepackage{fancyhdr}
\usepackage{verbatim}

\newcommand\ltap{\
  \raise.3ex\hbox{$<$\kern-.75em\lower1ex\hbox{$\sim$}}\ }
\newcommand\gtap{\
  \raise.3ex\hbox{$>$\kern-.75em\lower1ex\hbox{$\sim$}}\ }

\newcommand\simge{\mathrel{%
   \rlap{\raise 0.511ex \hbox{$>$}}{\lower 0.511ex \hbox{$\sim$}}}}
\newcommand\simle{\mathrel{
   \rlap{\raise 0.511ex \hbox{$<$}}{\lower 0.511ex \hbox{$\sim$}}}}

\newcommand{\slashchar}[1]%
        {\kern .25em\raise.18ex\hbox{$/$}\kern-.75em #1}
\def\lsim{\mathrel{\raise.3ex\hbox{$<$\kern-.75em\lower1ex\hbox{$\sim$}}}}
\def\gsim{\mathrel{\raise.3ex\hbox{$>$\kern-.75em\lower1ex\hbox{$\sim$}}}}
\newcommand{\bs}{\boldsymbol}
\newcommand{\Tr}{{\rm Tr}}
\newcommand\CA{{\cal A}}

\newcommand\CG{{\cal G}}

\newcommand\CL{{\cal L}}
\newcommand\CM{{\cal M}}
\newcommand\CN{{\cal N}}
\newcommand\CO{{\cal O}}
\newcommand\CP{{\cal P}}

\newcommand\be{\begin{equation}}
\newcommand\ee{\end{equation}}
\newcommand\bea{\begin{eqnarray}}
\newcommand\eea{\end{eqnarray}}
\newcommand\ba{\begin{array}}
\newcommand\ea{\end{array}}
\newcommand\nn{\nonumber}

\newcommand{\half}{\ensuremath{\frac{1}{2}}}
\newcommand{\thalf}{\textstyle{\frac{1}{2}}}
\newcommand{\third}{\ensuremath{\frac{1}{3}}}

\newcommand{\fourth}{\ensuremath{\frac{1}{4}}}
\newcommand{\tfourth}{\textstyle{\frac{1}{4}}}

\newcommand{\thw}{\ensuremath{\theta_W}}

\newcommand\dagg{\dagger}
\newcommand\ts{\thinspace}
\newcommand\ra{\rightarrow}

\newcommand\Lra{\Longrightarrow}
\newcommand\longra{\longrightarrow}
\newcommand\leftra{\leftrightarrow}

\newcommand\olra{\overleftrightarrow}
\newcommand\ol{\bar}
\newcommand\mev{{\rm MeV}}
\newcommand\gev{{\rm GeV}}
\newcommand\tev{{\rm TeV}}

\newcommand\ifb{{\rm fb}^{-1}}

\newcommand\shat{\hat s}

\newcommand\condtc{{\langle \ol T T \rangle}_{TC}}
\newcommand\condetc{{\langle \ol T T \rangle}_{ETC}}

\newcommand\ellm{\ell^-}

\newcommand\ellp{\ell^+}

\newcommand\suc{SU(3)_C}
\newcommand\Ntc{N_{TC}}
\newcommand\sutc{SU(N_{TC})}

\newcommand\Metc{M_{ETC}}

\newcommand\Ltc{\Lambda_{TC}}
\newcommand\Leff{{\cal L}_{\rm eff}}
\newcommand\Lsig{{\cal L}_{\Sigma}}
\newcommand\LFF{{\cal L}_{\rm gauge}}
\newcommand\LWZW{{\cal L}_{\rm WZW}}
\newcommand\Lff{{\cal L}_{\bar f f}}
\newcommand\Lpifbf{{\cal L}_{\tpi \bar f f}}
\newcommand\grpp{g_{\rho_T\pi_T\pi_T}}

\newcommand\tom{\omega_{T}}
\newcommand\tro{\rho_{T}}
\newcommand\atro{\alpha_{\rho_T}}

\newcommand\ta{a_T}

\newcommand\tap{a_T^+}
\newcommand\tam{a_T^-}
\newcommand\tapm{a_T^\pm}

\newcommand\taz{a_T^0}

\newcommand\tropm{\rho_{T}^\pm}

\newcommand\trop{\rho_{T}^+}
\newcommand\trom{\rho_{T}^-}
\newcommand\troz{\rho_{T}^0}

\newcommand\tpi{\pi_T}
\newcommand\tpipm{\pi_T^\pm}
\newcommand\tpimp{\pi_T^\mp}
\newcommand\tpip{\pi_T^+}
\newcommand\tpim{\pi_T^-}
\newcommand\tpiz{\pi_T^0}
\newcommand\tpipr{\pi_T^{0 \prime}}

\newcommand\chipr{\chi^{\ts \prime}}
\newenvironment{changemargin}[2]{\begin{list}{}{
        \setlength{\topsep}{0pt}\setlength{\leftmargin}{0pt}
        \setlength{\rightmargin}{0pt}
        \setlength{\listparindent}{\parindent}
        \setlength{\itemindent}{\parindent}
        \setlength{\parsep}{0pt plus 1pt}
        \addtolength{\leftmargin}{#1}\addtolength{\rightmargin}{#2}
        }\item }{\end{list}}
\begin{document}
\title{
\vskip -15mm
\begin{flushright}
\vskip -15mm
\vskip 5mm
\end{flushright}
{\Large{\bf An Effective Lagrangian for Low-Scale Technicolor}}\\
} \author{
  {\large Kenneth Lane$^{1,2}$\thanks{lane@physics.bu.edu} \, and
    Adam Martin$^{3}$\thanks{adam.martin@yale.edu}}\\
  {\large $^{1}$Department of Physics, Boston University}\\
  {\large Boston, Massachusetts 02215, USA}\\
  {\large $^{2}$LAPTH, Universit\'e de Savoie, CNRS} \\
  {\large B.P.~110, F-74941, Annecy-le-Vieux, France}\\
  {\large $^{3}$Department of Physics, Sloane Laboratory, Yale University}\\
  {\large New Haven, Connecticut 06520, USA}\\
}
\maketitle

\begin{abstract}
  We present an effective Lagrangian for low-scale technicolor. It describes
  the interactions at energies $\simle M_{\tro}$ of the lowest-lying bound
  states of the lightest technifermion doublet --- the spin-one
  $\tro,\tom,\ta,f_T$ and the corresponding technipions $\tpi$. This
  Lagrangian is intended to put on firmer ground the technicolor straw-man
  phenomenology used for collider searches of low-scale technicolor. The
  technivectors are described using the hidden local symmetry (HLS) formalism
  of Bando, {\em et al.} The Lagrangian is based on $SU(2)\otimes U(1)\otimes
  U(2)_L \otimes U(2)_R$, where $SU(2)\otimes U(1)$ is the electroweak gauge
  group and $U(2)_L \otimes U(2)_R$ is the HLS gauge group. Special attention
  is paid to the higher-derivative standard HLS and Wess-Zumino-Witten
  interactions needed to describe radiative and other decays of $\ta$ and
  $\tro/\tom$, respectively.

\end{abstract}


\newpage

\section*{I. Introduction and Motivation}

This paper is devoted to constructing an effective Lagrangian for low-scale
technicolor and discussing its main predictions for the Large Hadron
Collider. To begin, we believe that technicolor, if it describes electroweak
symmetry breaking, must have technihadron states at a low scale --- at just a
few hundred GeV. This section explains why we think this is so and our plan
for the paper.

Technicolor~\cite{Weinberg:1979bn,Susskind:1978ms,Lane:2002wv,Hill:2002ap}
was invented to provide a natural and consistent quantum-field-theoretic
description of electroweak (EW) symmetry breaking --- {\em without}
elementary scalar fields.  Extended technicolor~\cite{Eichten:1979ah,
  Dimopoulos:1979es} was invented to complete that natural description by
including quark and lepton flavors and their symmetry breaking as
interactions of fermions and gauge bosons alone.  At the outset, ETC was
recognized to have a problem with flavor-changing neutral current
interactions, especially those that induce $K^0$--$\bar K^0$ mixing.  The
problem is that very high ETC scales, of $\CO(1000\,\tev)$, are required to
suppress these interactions to an acceptable level while --- making plausible
QCD-based assumptions for the magnitudes of technifermion condensates
$\langle \bar T T \rangle$ --- ETC masses this large imply quark and lepton
masses of at most a few~MeV.  Walking
technicolor~\cite{Holdom:1981rm,Appelquist:1986an,Yamawaki:1986zg,
  Akiba:1986rr} was invented to cure this FCNC problem. The cure is that the
QCD-based assumptions do not apply to technicolor after all. In walking TC
the gauge coupling decreases very slowly, staying large for 100s, perhaps
1000s, of TeV and remaining near its critical value for spontaneous chiral
symmetry breaking. Then, the $\bar T T$ anomalous dimension $\gamma_m \simeq
1$ over this large energy range~\cite{Cohen:1988sq}, so that $\condetc \gg
\condtc$ and reasonable fermion masses result.\footnote{Except for the top
  quark, which needs an interaction such as topcolor to explain its large
  mass~\cite{Hill:1994hp}.}

Because it implies strong dynamics very different from QCD, a walking TC
gauge coupling may solve another problem, one of TC, not ETC. This is the
apparent conflict with precision electroweak measurements, especially with
the value of the $S$-parameter~\cite{Peskin:1990zt,Golden:1990ig,
  Holdom:1990tc,Altarelli:1991fk} extracted from these measurements. The
technicolor contribution to $S$ is defined in terms of polarization functions
of the technifermion electroweak currents and their spectral representation
by
\be\label{eq:sparameter}
S = 16\pi \frac{d}{d q^2} \left[\Pi_{33} (q^2) - \Pi_{3Q}(q^2) \right
]_{q^2=0}
= 4\pi \int\frac{dm^2}{m^4}\left[\sigma^3_V(m^2) - \sigma^3_A(m^2)\right]\,.
\ee
Here, for $N_D$ electroweak doublets of technifermions, the currents are
$j_{L\mu}^3 = \sum_{i=1}^{N_D} \half \bar T_{Li}\gamma_\mu\tau_3 T_{Li}$,
etc., and the $\sigma^3_{V,A}$ are vector and axial vector spectral functions
for the isovector currents $j^3_{\mu,\,5\mu}$. Experimentally, the
$S$-parameter is consistent with zero or slightly
negative~\cite{Amsler:2008zzb}. If the spectral functions in
Eq.~(\ref{eq:sparameter}) can be represented as a sum over a tower of narrow
isovector $\tro$ and $\ta$ resonances, with masses $M_{\rho_{T_i}}$,
$M_{a_{T_i}}$ and couplings $1/g_{\rho_{T_i}}$, $1/g_{a_{T_i}}$ to
$j^3_{\mu,\,5\mu}$ so that, e.g., $\sigma^3_V(m^2) \simeq \sum_{i=1}^{N_D}
\sum_{\rm{tower}} M_{\rho_{T_i}}^4/g_{\rho_{T_i}}^2 \delta(m^2 -
M_{\rho_{T_i}}^2)$, the $S$-parameter is given by
\be\label{eq:Snarrow}
S \simeq 4\pi\sum_{i=1}^{N_D} \sum_{\rm{tower}} \left[\frac{1}{g_{\rho_{T_i}}^2}
    - \frac{1}{g_{a_{T_i}}^2} \right] \,.
\ee
%
 
The usual assumptions made to estimate the TC contribution to $S$ are based
on analogy with the way QCD actually works. These assumptions are invalid in
walking technicolor~\cite{Lane:1993wz,Lane:1994pg}. In particular, in QCD the
lowest lying $\rho$ and $a_1$ saturate the integrals appearing in Weinberg's
spectral function sum rules~\cite{Weinberg:1967kj,Bernard:1975cd}.  Then,
$\Pi_{33} - \Pi_{3Q}$ falls off like $1/q^4$ for $q^2 \simge 1\,\gev^2$, and
the spectral integrals for the sum rules and $S$ converge very rapidly. This
``vector meson dominance'' of the spectral integrals is related to the
precocious onset of asymptotic freedom in QCD.  The $1/q^4$ behavior is
consistent (up to logs) with the operator product expansion. The leading OPE
term for $\Pi_{33} - \Pi_{3Q}$ is essentially $\langle\ol T T \ol T
T\rangle_q/q^4$, and it dominates above $\sim 1\,\gev$.  Here, the
$q$-subscript indicates the scale at which the operator is renormalized. In
walking TC, however, $\langle\ol T T \ol T T\rangle_q \sim (q^2/\Ltc^2)
\langle\ol T T \ol T T\rangle_{\Ltc}$ for $q$ below the scale at which
asymptotic freedom finally sets in. To account for this in terms of spin-one
technihadrons, the tower of $\tro$ and $\ta$ must extend to very high energy
and contribute substantially to the spectral function sum rules and to $S$.
Lacking experimental knowledge of these states, and even whether a tower of
states is the proper description of the spectral functions, it is at least as
difficult to estimate $S$ reliably for TC as it would have been for QCD
before the $\rho$ and $a_1$ were discovered.  Undaunted, some theorists in
the past decade suggested how walking (or near-conformal) dynamics might
solve the $S$-parameter problem; see, e.g., Refs.~\cite{Appelquist:1998xf,
  Appelquist:1999dq, Hirn:2006nt, Hirn:2006wg,Foadi:2007ue}.  These
proposals, in their simplest realization, amount to there being near equality
of the partner $\tro$--$\ta$ masses and couplings to the weak currents. For
related work, see~\cite{Knecht:1997ts, Casalbuoni:1995qt, Eichten:2007sx}.
Thus, $S$ may be small, and even negative if
\be\label{eq:equal}
M_{a_{T_i}} \cong M_{\rho_{T_i}} \quad {\rm and} \quad
g_{a_{T_i}} \cong g_{\rho_{T_i}}\,.
\ee
This is an interesting and reasonable assumption, and it may be more
plausible than requiring large cancellations among the many TC contributions
to $S$. But, just how walking technicolor produces this result is a knotty
theoretical problem.\footnote{It is possible that some sort of duality
  connecting walking TC to a weakly coupled theory, perhaps in higher
  dimensions, can explain this. See e.g., Refs.~\cite{Hirn:2006nt,
    Hirn:2006wg}. Contrarily, see e.g. Ref.~\cite{Agashe:2003zs}}. We shall
see that, depending on the relative size of couplings in our effective
Lagrangian, there can be tension between Eq.~(\ref{eq:equal}) and the
phenomenology of low-scale technicolor.

A walking TC gauge coupling with $\gamma_m \simeq 1$ for a large energy range
occurs if the critical coupling for chiral symmetry breaking lies just {\em
  below} a value at which there is an infrared fixed
point~\cite{Lane:1991qh,Appelquist:1997fp}. This requires a large number of
technifermions. That may be achieved by having $N_D \gg 1$ doublets in the
fundamental representation $\bs{N}_{TC}$ of the TC gauge group, $SU(N_{TC})$,
or by having a few doublets in higher-dimensional
representations~\cite{Lane:1989ej,Dietrich:2005wk}. In the latter case, the
constraints on ETC representations~\cite{Eichten:1979ah} almost always imply
other technifermions in the fundamental representation as well. In either
case, then, there generally are technifermions whose technipion ($\tpi$)
bound states have a decay constant $F_1^2 \ll F_\pi^2 = (246\,\gev)^2$. This
low scale implies there also are technihadrons $\tro$, $\tom$, $\ta$, etc.
with masses well below a TeV. We refer to this situation as low-scale
technicolor (LSTC)~\cite{Lane:1989ej, Eichten:1996dx,Eichten:1997yq}. While,
in the past, we preferred to assume the alternative of many TC fundamentals,
the effective Lagrangian we present below is applicable to either
situation.\footnote{Two other walking-TC scenarios have been proposed and
  these need not have low-mass technifermions.
  Ref.~\cite{Christensen:2005cb} employed electroweak singlet technifermions
  to make the TC coupling walk.  Ref.~\cite{Evans:2005pu} considered an
  $SU(2)_{TC}$ with technifermions in the vector representation and a
  TC-singlet lepton doublet.}

We stress two important consequences of this picture of walking TC. First, to
restate what we just said, $N_D > 1$ technifermion doublets implies the
existence of physical technipions, some of which couple to the lightest
technivector mesons. Second, since $M_{\tpi}^2 \propto \langle\ol T T \ol T
T\rangle_{\Metc}$, walking TC enhances the masses of technipions much more
than it does other technihadron masses. Thus, it is very likely that the
lightest $M_{\tro} < 2M_{\tpi}$ and that the two and three-$\tpi$ decay
channels of the light technivectors are closed~\cite{Lane:1989ej}. This
further implies that these technivectors are {\em very} narrow, a few~GeV or
less, because their decay rates are suppressed by phase space and/or small
couplings (see below). Technipions are a distinctive feature of LSTC and
finding them in the decays of technivectors is an important way of
distinguishing it from other scenarios of dynamical electroweak symmetry
breaking, such as Higgsless models in five dimensions~\cite{Csaki:2003dt,
  Csaki:2003zu, Agashe:2003zs, Cacciapaglia:2004rb}, deconstructed
models~\cite{SekharChivukula:2004mu, Chivukula:2005bn}, a walking TC model
with $\Ntc = 2$ and just one doublet of
technifermions~\cite{Sannino:2004qp,Dietrich:2005wk,Foadi:2007ue}, and the
BESS and DBESS models~\cite{Casalbuoni:1988xm,Casalbuoni:1995qt}.

A simple phenomenology of LSTC is provided by the Technicolor Straw-Man Model
(TCSM)~\cite{Lane:1999uh,Lane:2002sm,Eichten:2007sx}. The TCSM's ground rules
and major parameters are these:

\begin{enumerate}
  
\item The lightest doublet of technifermions $(T_U,T_D)$ are color-$\suc$
  singlets.\footnote{Colored technifermions get a substantial contribution to
    their mass from $\suc$ gluon exchange. We also assume implicitly that, in
    the case of $N_D$ fundamentals, ETC interactions split the doublets
    substantially.}
  
\item The decay constant of the lightest doublet's technipions is $F_1 =
  (F_\pi = 246\,\gev)\cdot \sin\chi$. In the case of $N_D$ fundamentals,
  $\sin^2\chi \cong 1/N_D \ll 1$. In the case of two-scale TC, $F_\pi =
  \sqrt{F_1^2 + F_2^2} = 246\,\gev$ with $F_1^2/F_2^2 \cong \tan^2\chi \ll
  1$.

\item The isospin breaking of $(T_U,T_D)$ is small. Their electric charges
  are $Q_U$ and $Q_D = Q_U - 1$. In Refs.~\cite{Lane:2002sm,
    Eichten:2007sx} the rates for several decay modes of the technivectors to
  transversely-polarized electroweak gauge bosons
  ($\gamma,W_\perp^\pm,Z_\perp^0$) plus a technipion or longitudinal weak
  boson ($W_L^{\pm,0} \equiv W_L^\pm,Z_L^0$) and for decays to a
  fermion-antifermion pair depend sensitively on $Q_U+Q_D$.
  
\item The lightest technihadrons are the pseudoscalars $\pi_{T1}^{\pm,0}(I =
  1)$, $\pi_{T1}^{0\prime}(I=0)$ and the vectors $\tro^{\pm,0}(I=1)$,
  $\tom(I=0)$ and axial vectors $a_T^{\pm,0}(I=1)$, $f_T(I=0)$. Isospin
  symmetry and quark-model experience strongly suggest $M_{\tro} \cong
  M_{\tom}$ and $M_{\ta} \cong M_{f_T}$.
  
\item Since $W_L^{\pm,0}$ are superpositions of all the isovector
  technipions, the $\pi_{T1}$ are not mass eigenstates. This is parameterized
  in the TCSM as a simple two-state admixture of $W_L$ and the lightest
  mass-eigenstate $\tpi$:
\be\label{eq:pistates}
 \vert\pi_{T1}\rangle = \sin\chi \ts \vert
W_L\rangle + \cos\chi \ts \vert\tpi\rangle\ts.
\ee
Thus, technivector decays involving $W_L$, while nominally, strong
interactions, are suppressed by powers of $\sin\chi$. In a similar way,
$\vert\pi_{T1}^{0 \prime} \rangle = \cos\chipr \ts \vert\tpipr\rangle\ +
\cdots$, where $\tpipr$ is the lightest isoscalar technipion, $\chipr$ is
another mixing angle, and the ellipsis refer to other isoscalar bound states
of technifermions needed to eliminate the two-technigluon anomaly from the
$\pi_{T1}^{0 \prime}$ chiral current. It is unclear whether $\tpiz$ and
$\tpipr$ will be approximately degenerate as $\tro$ and $\tom$ are. While
they both contain the lightest $\ol T T$ as constituents, $\tpipr$ must
contain other heavier technifermions because of the anomaly cancellation.

\item The lightest technihadrons, $\tpi$, $\tro$, $\tom$ and $\ta$, may be
  studied {\em in isolation}, without significant mixing or other
  interference from higher-mass states. This is the most important of the
  TCSM's assumptions. It is made to avoid a forest of parameters, and it is
  in accord with the ``simplicity principle'' for our effective Lagrangian,
  discussed below. In the absence of actual data on technihadrons, there is
  no way to know its validity.
  
\item In addition to these technihadrons and $W_L^\pm$, $Z^0_L$, the TCSM
  involves the transversely-polarized $\gamma$, $W^\pm_\perp$ and
  $Z^0_\perp$. The principal production process of the technivector mesons at
  hadron and lepton colliders is Drell-Yan, e.g, $\ol q q \ra \gamma,Z^0 \ra
  \troz, \tom, \taz \ra X$. This gives strikingly narrow $s$-channel
  resonances at $M_X = M_{\troz,\tom,\ta}$ {\em if} $M_X$ can be
  reconstructed.
  
\item Technipion decays are mediated by ETC interactions and are therefore
  expected to be Higgs-like, i.e., $\tpi$ preferentially decay to the
  heaviest fermion pairs they can. There are two exceptions. Something like
  topcolor-assisted technicolor~\cite{Hill:1994hp} is required to give the
  top quark its large mass. Then, the coupling of $\tpi$ to top quarks is not
  proportional to $m_t$, but more likely to $\CO(m_b)$~\cite{Hill:1994hp}.
  We shall take this into account in Sec.~V (see Eq.~(\ref{eq:Lpiff})).
  Second, the two-gluon decay mode of $\tpipr$ can be appreciable which would
  make it difficult to discover it at a hadron collider. In this paper we
  shall assume that the $\tpipr$ is heavier than the other LSTC hadrons. Then
  it is not interesting phenomenologically and we shall not study the details
  of its interactions.

\end{enumerate}

This TCSM phenomenology was tested at LEP (see, e.g.,
Refs.~\cite{Abdallah:2001ft,Schael:2004tq}) and the
Tevatron~\cite{Abazov:2006iq,CDFa,Nagai:2008xq} for certain generic values of
the parameters. So far there is no compelling evidence for TC, but there are
also no significant restrictions on the masses and couplings commonly used in
the TCSM search analyses carried out so far ($M_{\tro} \simge
225$--$250\,\gev$, $M_{\tpi} \simge 125$--$145\,\gev$, $\sin\chi = 1/3$ and
$Q_U \simeq 1$). On the other hand, the more general idea of LSTC makes
little sense if the limit on $M_{\tro}$ is pushed past 600--$700\,\gev$.
Therefore, we believe that the LHC can discover it or certainly rule it
out~\cite{Brooijmans:2008se}. If LSTC were found at the LHC, it would be a
field day for a linear collider such as the ILC or CLIC with $\sqrt{s} \simeq
M_{\tro,\tom,\ta}$.  Such a collider may be able to separate the closely
spaced $\troz$ and $\tom$ and, perhaps, $\taz$ resonances. Furthermore,
precision measurements, essentially free of background, of the rates and
angular distributions of these states' decays into gauge boson and
$\ellp\ellm$ pairs could yield valuable information on LSTC masses and
couplings.

The TCSM described above was incorporated into {\sc
  Pythia}~\cite{Sjostrand:2006za} and used in the recent CDF
study~\cite{Nagai:2008xq}. Nowadays, however, many physicists prefer the
versatility of programs such as CalcHEP~\footnote{{\tt
    http://theory.sinp.msu.ru/~pukhov/calchep.html}}, MadGraph~\footnote{{\tt
    madgraph.hep.uiuc.edu/}} and SHERPA~\footnote{{\tt
    http://projects.hepforge.org/sherpa/dokuwiki/doku.php}} to generate new
physics signal and background events at the parton level. CalcHEP {\em et
  al.} require inputting a set of Feynman rules, consistent with all relevant
gauge and global symmetries. From these, they generate scattering amplitudes
that can be interfaced with such programs as {\sc Pythia} and
HERWIG~\footnote{{\tt
    http://hepwww.rl.ac.uk/theory/seymour/herwig/herwig65.html}} for decays
and hadronization.

The Feynman rules, of course, require a Lagrangian. So far, however, a
Lagrangian has not been written down for the TCSM or any other variant of
LSTC. This is because of the way it was formulated and implemented in {\sc
  Pythia}. To guarantee a massless photon pole, {\em kinetic} mixing was used
in inverse propagator matrices describing the coupling between gauge and
technivector bosons. These large matrices must then be inverted at each value
of $\shat$ for use in the amplitudes for processes enhanced by $\tro$, $\tom$
and $\ta$ poles such as $\bar q q' \ra W^\pm \tpi$ and $W^\pm Z^0$.  Another
feature difficult to include in a Lagrangian is the way the TCSM described
production of longitudinal weak bosons. Amplitudes involving $W_L$ treated
them as spinless particles which, although not a bad approximation at LHC
energies, is exact only when $\sqrt{\shat} \gg M_W$. This treatment makes it
especially difficult in the TCSM to discuss properly, e.g, the
$\tro$-enhancement of $W_L^+ W_L^-$ in $e^+e^- \ra W^+ W^-$.

The purpose of this paper is to provide an effective Lagrangian, $\Leff$, for
low-scale technicolor. It includes all the LSTC states listed above plus the
quarks and leptons. This $\Leff$ is ``effective'' not only in being valid
just in the energy region in which one can consider the lowest-lying
technihadrons in isolation. In LSTC, typical momenta are of the order of the
scale --- which we shall call $F_1$ --- which ``suppresses'' higher
derivative terms, so there seems no systematic way to limit the terms
included. Much the same is true of an effective Lagrangian for QCD if the
$\rho$ and $a_1$ mesons are included. We shall adhere to a ``principle of
simplicity'': we keep only the lowest-dimension operators sufficient to
describe the phenomenologically important processes of LSTC. Thus, as in the
{\sc Pythia} implementation of the TCSM, we strive to minimize the number of
adjustable parameters in $\Leff$.

We adopt the hidden local symmetry (HLS) formalism of Bando, {\it et
  al.}~\cite{Bando:1984ej, Bando:1987br} to describe the technivector mesons,
electroweak bosons and technipions. This method guarantees that the photon is
massless and the electromagnetic current conserved. The ``naive'' form of the
HLS Lagrangian, $\Lsig$, in which terms with no more than two covariant
derivatives are kept, also guarantees that production of longitudinal
electroweak bosons via annihilation of massless fermions is well-behaved at
all energies {\em in tree approximation.}\footnote{The reason for this is
  that, in the absence of $\Lsig$, the gauge structure of the Lagrangian,
  including HLS interactions, ensures good high-energy behavior of gauge
  boson production.  Turning on the naive $\Lsig$ merely mixes the gauge
  bosons while leaving the EW symmetry structure of triple and quartic gauge
  boson interactions unaltered. This argument will be modified when we add
  higher derivative interactions to $\Lsig$.} This is important, because many
of the most experimentally accessible LSTC processes at colliders involve
production of one or more $W_L$ in the final state. Elastic $W_L W_L$
scattering still behaves at high energy as it does in the standard model
without a Higgs boson, i.e., the amplitude $\sim s/F_\pi^2$ at large cm
energy~$s$. Of course, this violation of perturbative unitarity signals the
strong interactions of the underlying technicolor theory.

Unfortunately, the naive HLS formalism is too restrictive. Its
two-covariant-derivative
structure and its symmetries imply relations for interaction operators which
are untrue for bound states such as $\tro$ (see, e.g,
Refs.~\cite{Bando:1985rf, Bando:1987br, Zerwekh:2001uq,
  SekharChivukula:2001gv}). Important LSTC processes, such as $\tapm \ra
\gamma \tpipm$, $\gamma W^\pm$ and $\tom,\,\tro \ra \gamma \tpiz$, $\gamma
Z^0$ do not occur in the Lagrangian. For the radiative and other $\ta$
decays, we shall apply our ``simplicity principle'' to choose one particular
four-derivative operator of many possible ones. One would expect this
operator to spoil the high energy behavior of amplitudes to which it
contributes. As we will show in Sec.~II, while amplitudes involving only
standard model (SM) particles may be modified by this new term, their
large-$s$ behavior is unaltered.

The absence of $\tro$ and $\tom$ radiative decays from the naive Lagrangian
is more serious. As in QCD, it happens because the Lagrangian has a parity
symmetry not present in the underlying theory. And, as in QCD, the remedy is
found in Wess-Zumino-Witten (WZW) terms~\cite{Wess:1971yu, Witten:1983tw}.
They implement the effects of anomalously nonconserved symmetries of the
high-energy theory --- in QCD, the Adler-Bardeen-Jackiw anomaly. In our case,
the question of the anomalies of the high-energy theory is even more subtle.
Partly, this is because the HLS Lagrangian seems to require a more extensive
set of fermions in the underlying theory than just the lightest doublet
$(T_U, T_D)$, so the anomalies in question are less obvious. In addition,
there is nothing to cancel the anomalies of the HLS gauge interaction so,
unlike the LSTC theory it is supposed to represent, it is truly
nonrenormalizable. A somewhat similar problem was considered by Harvey, Hill
and Hill~\cite{Harvey:2007ca} (extending and improving earlier work of
Kaymakcalan, Rajeev and Schechter~\cite{Kaymakcalan:1983qq}). They
constructed a gauge-invariant WZW interaction for the standard model in the
presence of $\rho$ and $a_1$, which they treated as background fields. An
obstacle for us was determining how to apply Ref.~\cite{Harvey:2007ca} to our
nonrenormalizable theory, in which the HLS fields are dynamical and mix with
the electroweak ones. To our knowledge, this has not been done previously for
a theory with anomaly-free, renormalizable gauge symmetries and anomalous
hidden local symmetries involving vector {\em and} axial vector mesons.


The HLS formalism has also been used in BESS models~\cite{Casalbuoni:1988xm,
  Casalbuoni:1995qt}, a minimal model of walking
technicolor~\cite{Foadi:2007ue}, and in deconstructed
versions~\cite{SekharChivukula:2004mu,Chivukula:2005bn,Chivukula:2005xm,
  Chivukula:2005ji,Foadi:2003xa} of five-dimensional Higgsless
models~\cite{Csaki:2003dt, Csaki:2003zu, Agashe:2003zs, Cacciapaglia:2004rb}.
However, these papers did not include higher-derivative interactions needed
for $\ta$ decays nor the WZW interactions for $\tro$ and $\tom$ decays.

The rest of this paper is organized as follows: Sec.~II specifies the
symmetries and gauge and Goldstone fields used to construct $\Leff$. Then, we
use LSTC dynamics (and phenomenology) to motivate the two-derivative terms we
allow in the naive $\Leff$. The resulting Lagrangian is similar, but not
identical, to those used in Refs.~\cite{Bando:1985rf} for QCD
and~\cite{Casalbuoni:1995qt} for strong electroweak symmetry breaking. We
differ from them in that we included the $U(1)_Y$ gauge boson and its
couplings to the technihadrons consistent with arbitrary $(T_U,T_D)$ charges
$Q_U$ and $Q_D = Q_U - 1$. Also, as we emphasized, other treatments of strong
electroweak symmetry breaking do not include technipions; we expect them to
occur in any realistic low-scale technicolor model. Finally, we added an
interaction to describe $a_T \ra \gamma \tpi$ and $\gamma W/Z$. A similar
interaction occurs in Ref.~\cite{Bando:1985rf}. In Sec.~III we transform to
the unitary gauge and present the vector boson mass matrices, eigenvalues and
eigenstates. The connections with the masses and mixings of the electroweak
and technivector bosons in the TCSM~\cite{Lane:2002sm, Eichten:2007sx} are
discussed. We describe the shifting of gauge fields necessary to eliminate
mixed gauge-technipion kinetic terms. The WZW interaction needed at low
energy to describe certain important technivector decays is treated in
Sec.~IV. As a test of the prescription we use to determine it, we show that it
produces the expected form for $\tpiz \ra \gamma\gamma.$ Technipion masses
and their couplings to quarks and leptons are given in Sec.~V.

In Sec.~VI we compare the predictions of our $\Leff$ with the TCSM
phenomenology outlined above for the technihadron decay amplitudes that are
important at the Tevatron and LHC. There is, in fact, no {\em a priori}
guarantee that our $\Leff$ reproduces the TCSM because, as we noted above, it
is not clear that the two have the same underlying technicolor theory.
Nevertheless, we find that they agree. In particular, terms in $\Leff$
related by the replacement of $\tpi^{\pm,0}$ by $W_L^{\pm,0}$ stand in the
ratio $\cos\chi:\sin\chi$, and amplitudes for processes such as
$\tro^{\pm,0} \ra \gamma \tpi^{\pm,0}$ and $\tom \ra \gamma \tpiz$ differ
only by simple valence-quark-model-like ``Clebsch'' factors. Indeed,
requiring that $\Leff$ reproduce the TCSM in this way has been a valuable
check on our calculations.

In Sec.~VII we use $\Leff$ to calculate the low-scale technihadrons'
contributions to the precision electroweak parameters $S$, $T$, $W$ and
$Y$~\cite{Peskin:1990zt, Maksymyk:1993zm,Barbieri:2004qk, Marandella:2005wd}
in tree approximation.  We see that, thanks to the higher-derivative term
added to account for $\ta$ decays, it is possible to make this contribution
to $S$ small. The matter of this term's contribution to TCSM phenomenology is
under investigation and will be the subject of a future paper. Finally, some
projects for future study are described in Sec.~VIII.

Three appendices are attached. Appendix~A summarizes the TCSM predictions
amplitudes for technivector decay to a pair of technipions and/or electroweak
bosons and is useful for comparison with the implications of $\Leff$.
Appendix~B contains the eigenvectors for the mass-eigenstate gauge bosons and
the coefficients $\zeta$ in the shifts of the gauge fields, $g_G G^\alpha_\mu
\ra g_G G^\alpha_\mu + \partial_\mu \zeta_{G^\alpha}/F_1$, needed to rectify
their kinetic energy terms.  Appendix~C is a list of $\Leff$'s adjustable
parameters and suggested defaults.

\section*{II. LSTC Symmetries and the Effective Lagrangian}

 \begin{figure}[!t]
   \begin{center}
     \vspace*{-0.5in}
     \includegraphics[width=7.00in, height = 3.00in, angle=0]
     {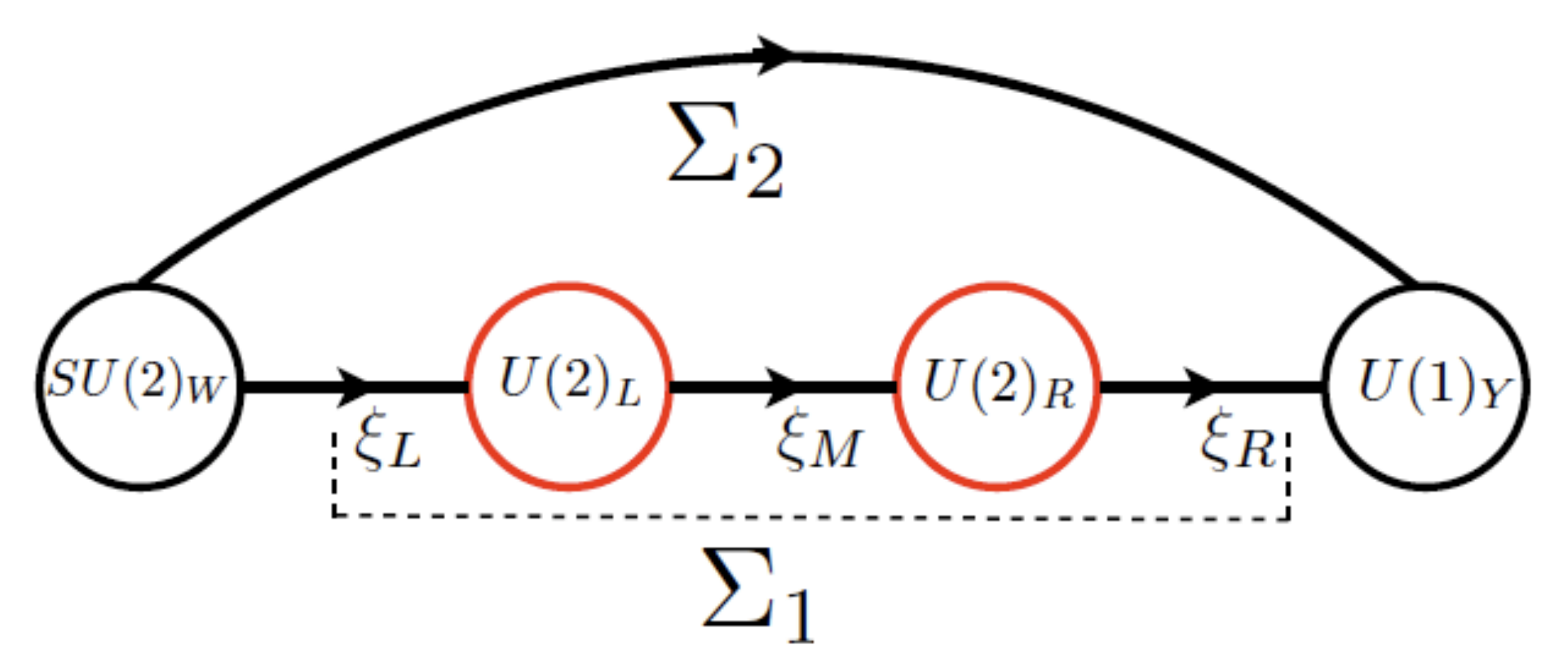}
     \caption{Moose diagram for the LSTC model defined by
       Eqs.~(\ref{eq:sigmatransforms}), with $y_1 = 0$.}
     \label{moose}
   \end{center}
 \end{figure}

 Our Lagrangian is based on the hidden local symmetry
 formalism~\cite{Bando:1984ej,Bando:1987br} with gauge group $\CG = SU(2)_W
 \otimes U(1)_Y \otimes U(2)_L \otimes U(2)_R$. The first two groups are the
 standard electroweak gauge symmetries, with primordial couplings $g$ and
 $g'$ and gauge bosons $\bs{W} = (W^1,W^2,W^3)$ and $B$, respectively. The
 latter two are the ``hidden local symmetry'' groups. We use $U(2)_{L,R}$
 instead of $SU(2)_{L,R}$ for the HLS groups because we expect the isoscalar
 $\tom$ to be important phenomenologically. Furthermore, radiative decays of
 $\troz$ and $\tom$ to the same final state differ only in a factor of $Q_U +
 Q_D$ versus $Q_U - Q_D = 1$. Thus, they can in principle tell us about
 technifermion charges. We assume that the underlying TC interactions are
 parity-invariant, so that their zeroth-order couplings are equal, $g_L = g_R
 = g_T$. The assumed equality of the $SU(2)_{L,R}$ and $U(1)_{L,R}$ couplings
 reflect the isospin symmetry of TC interactions and the expectation that
 $M_{\tro} \cong M_{\tom}$ and $M_{\ta} \cong M_{f_T}$. The gauge bosons
 $(\bs{L},L^0)$ and $(\bs{R},R^0)$ contain the primordial technivector
 mesons, $\bs{V},V_0,\bs{A},A_0 \cong \bs{\rho}_T,\tom,\bs{a}_T,f_T$. We
 shall see in Sec.~III that we can identify $g_T \simeq \sqrt{2}g_{\tro}$,
 where $g_{\tro}$ is the $\tro$ coupling to the isospin current (see
 Eq.~(\ref{eq:Snarrow})).
  
 To describe the lightest $\tpi$ and $\tpipr$, and to mock up the heavier TC
 states that contribute most to electroweak symmetry breaking (i.e., the
 isovector technipions of the other $N_D - 1$ technifermion doublets or the
 higher-scale states of a two-scale TC model), and to break all the gauge
 symmetries down to electromagnetic $U(1)$, we use nonlinear $\Sigma$-model
 fields $\Sigma_2$, $\xi_L$, $\xi_R$ and $\xi_M$. Under $\CG$-transformations
 (see the moose diagram in Fig.~\ref{moose}):
\bea\label{eq:sigmatransforms}
\Sigma_2 &\ra& U_W\Sigma_2U_Y^\dagg\nn \\
\xi_L &\ra& U_W U_Y \xi_L U_L^\dagg \nn \\
\xi_M &\ra& U_L \xi_M U_R^\dagg \nn \\
\xi_R &\ra& U_R \xi_R U_Y^\dagg\,.
\eea
Their covariant derivatives are
\bea\label{eq:covderivs}
D_\mu \Sigma_2 &=& \partial_\mu\Sigma_2 -ig \bs{t}\cdot\bs{W}_\mu \Sigma_2
+ ig' \Sigma_2 t_3 B_\mu  \nn \\
D_\mu \xi_L &=& \partial_\mu\xi_L -i(g\bs{t}\cdot\bs{W}_\mu +g' y_1 t_0
B_\mu) \xi_L + ig_T \xi_L\, t\cdot L_\mu\nn\\ 
D_\mu \xi_M &=& \partial_\mu\xi_M  -ig_T (t \cdot L_\mu\, \xi_M - \xi_M\,  t \cdot
R_\mu) \nn\\ 
D_\mu \xi_R &=& \partial_\mu\xi_R -ig_T t \cdot R_\mu\, \xi_R + ig' \xi_R (t_3
+ y_1 t_0)B_\mu \,,
\eea
where $t\cdot L_\mu = \sum_{\alpha=0}^3 t_\alpha L_\mu^\alpha$ and $\bs{t} =
{\half}\bs{\tau}$, $t_0 = {\half}\bs{1}$. The hypercharge $y_1 = Q_U + Q_D$
of the TCSM. The field $\Sigma_2$ contains the technipions that get absorbed
by the $W$ and $Z$ bosons. We represent them as an isotriplet of $F_2$-scale
Goldstone bosons, where $F_2 = F_\pi \cos\chi \gg F_1$, and $\chi$ was
introduced in Sec.~I.\footnote{Chivukula, {\em et
    al.}~\cite{Chivukula:2009ck} recently used an HLS construction with
  multiple scales and a parameter analogous to $\chi$ to discuss a Higgsless
  model with topcolor for top-quark mass generation.} It may be
parameterized as $\Sigma_2(x) = \exp{(2i\bs{t}\cdot\bs{\pi}_2(x)/F_2)}$. We
define $\Sigma_1 = \xi_L \xi_M \xi_R$. Then
\bea\label{eq:sigone}
\Sigma_1 &\ra& U_W \Sigma_1 U_Y^\dagg \nn\\
D_\mu \Sigma_1 &=& \partial_\mu\Sigma_1 -ig \bs{t}\cdot\bs{W}_\mu \Sigma_1
+ ig' \Sigma_1 t_3 B_\mu\,.
\eea

To construct an effective Lagrangian of manageable size, we first include
only two-derivative terms of the nonlinear fields. There is not much
justification for this in LSTC because the momenta of technivector decay
products are typically of order $F_1$. There are still $\CO(10)$ possible
$\Tr(|D_\mu\Sigma|^2)$ terms. We limit them by requiring that $g_T$-strength
interactions ({\em only!}) are consistent with the assumptions of their
underlying LSTC dynamics: In particular, they are isospin and
parity-invariant. To reduce their number, we employ the TCSM assumption that
the lowest-lying technihadrons {\em in isolation}. This implies that the
interactions we allow arise only from two-technifermion irreducible (i.e,
Zweig-allowed) graphs.\footnote{This eliminates many interactions, e.g.,
  $\Tr|D_\mu\Sigma_1\Sigma_2^\dagg|^2$ and $\Tr(D_\mu\Sigma_1\Sigma_1^\dagg)
  \Tr(D_\mu\Sigma_2\Sigma_2^\dagg)$.}. Then the {\em naive} version of the
nonlinear Lagrangian is
\bea\label{eq:Lsigzero}
\Lsig = && {\tfourth} F_2^2 \Tr|D_\mu\Sigma_2|^2 
      + {\tfourth} F_1^2 \Bigl\{a \Tr|D_\mu\Sigma_1|^2 + 
        b\Bigl[\Tr|D_\mu\xi_L|^2 + \Tr|D_\mu\xi_R|^2\Bigr] \nn\\
        && + c\, \Tr|D_\mu\xi_M|^2 
           + d\, \Tr(\xi_L^\dagg D_\mu \xi_L D_\mu \xi_M \xi_M^\dagg + 
                   \xi_R D_\mu \xi_R^\dagg D_\mu \xi_M^\dagg \xi_M)\Bigr\}
\eea
The couplings $a,b,c,d$ are nominally of order one in magnitude.

The interaction in Eq.~(\ref{eq:Lsigzero}) does not contain terms for the
phenomenologically important decays $\tapm \ra \gamma \tpipm$ (and, of
especial importance at the LHC, $\tapm \ra \gamma
W^\pm$~\cite{Brooijmans:2008se}). Gauge invariance and parity conservation
require these be mediated by terms of the form $F_{\mu\nu}(a_T)
F_{\mu\nu}(\gamma) \tpi$ and, so, we must include higher derivative terms in
$\Lsig$ to do the job. The same problem was faced for QCD in
Ref.~\cite{Bando:1987br}. Unlike those authors, we have no experimental input
to guide us, so we assume our ``simplicity principle'' and add just one
four-derivative term to $\Lsig$:
\bea\label{eq:Lsigone}
\Lsig = && {\tfourth} F_2^2 \Tr|D_\mu\Sigma_2|^2 
      + {\tfourth} F_1^2 \Bigl\{a \Tr|D_\mu\Sigma_1|^2 + 
        b\Bigl[\Tr|D_\mu\xi_L|^2 + \Tr|D_\mu\xi_R|^2\Bigr] \nn\\
        && + c\, \Tr|D_\mu\xi_M|^2 
           + d\, \Tr(\xi_L^\dagg D_\mu \xi_L D_\mu \xi_M \xi_M^\dagg + 
                   \xi_R D_\mu \xi_R^\dagg D_\mu \xi_M^\dagg
                   \xi_M)\nn\\
        && - \frac{if}{2g_T}\,\Tr(D_\mu\xi_M \xi_M^\dagg D_\nu\xi_M
        \xi_M^\dagg\, t\cdot L_{\mu\nu} + 
                    \xi_M^\dagg D_\mu\xi_M \xi_M^\dagg D_\nu\xi_M\,
                    t\cdot R_{\mu\nu})\Bigr\}\,.
\eea
As with the other constants, we expect $f = \CO(1)$. The normalization of the
$f$-term is chosen to make its contribution to $\tro \ra \tpi\tpi$ easy to
compare with that from other terms. As we shall see in Sec.~VI, the
decays $\ta \ra W/Z + \tpi$ and $\tapm \ra W^\pm Z^0$ also proceed through the
$f$-term. Several of these modes will be sought at the Tevatron and the
LHC~\cite{Brooijmans:2008se}.

One expects higher-derivative operators such as this $f$-term to spoil the
high energy behavior of standard-model processes. Fortunately, while the
$f$-term may modify their form and field structure, as we explain now, it
does not alter the dependence of $SM \ra SM$ amplitudes on the cm energy~$s$
at high energy: All mixing among the gauge bosons is induced by the
two-derivative terms in $\Lsig$. If there were no mixing, the f-term would
not contribute to any $SM \ra SM$ process, and the high energy behavior of
such amplitudes would be as in the standard model without a Higgs boson. In
particular, the amplitudes for massless fermion-antifermion annihilation to a
pair of longitudinally polarized gauge bosons would be constant at high
energy and the $W_L^+ W_L^- \ra W_L^+ W_L^-$ amplitude would grow linearly
with~$s$. Suppose we turn on the mixing between the primordial EW gauge
bosons, $W$ and $B$, and the $R,L = (V \pm A)/\sqrt{2}$ bosons in such a way
that all acquire the same mass. Then, the unitary transformation matrices
from the gauge basis to the mass basis are undefined, and nothing, including
the effect of the $f$-term on $SM \ra SM$ processes, can depend on them. In
this case then, the $f$-term still does not contribute to these processes.
Now, allowing different masses for the gauge bosons, it is clear that the
$f$-term contribution to $SM \ra SM$ amplitudes must involve differences of
gauge boson propagators, differences which vanish when the bosons are
degenerate.  This reduces the high-$s$ behavior of these amplitudes by one
power of~$s$ from naive power-counting, and so they have the large-$s$
dependence expected in the standard model (without a Higgs). In particular,
$f \bar f \ra WW,\, WZ \sim\,\, {\rm constant}$ and $W_L^+ W_L^- \ra W_L^+
W_L^- \sim s/F_\pi^2$ at large~$s$.

The effect of the $f$-term (as well as other terms in $\Lsig$) on these
standard-model processes and on triple gauge boson vertices remains to
be worked out.  These are under investigation and will be the subject of
future papers (see Sec. VIII).


Still, $\Lsig$ does not allow $\tpiz \ra \gamma\gamma$ and $\tro,\tom \ra
\gamma \tpi$. The reason for this is that the $g_T$-strength interactions in
$\Lsig$ are invariant under more than isospin and space inversion. Under
ordinary parity, $P$: ${\bs r} \ra -{\bs r}$, $t \ra t$, and $\xi_{L,R}
\leftra \xi_{R,L}^\dagg$, $\xi_M \leftra \xi_M^\dagg$, $\Sigma_i \leftra
\Sigma_i^\dagg$ and $R_\mu^\alpha \leftra (-1)^{(1+g_{\mu 0})} L_\mu^\alpha$.
Generalizing the discussion in Ref.~\cite{Witten:1983tw}, the strong
interactions in Eq.~(\ref{eq:Lsigzero}) are also invariant under $P_0$: ${\bs
  r} \ra -{\bs r}$, $t \ra t$, $(R,L)_\mu \ra (-1)^{1+g_{\mu 0}} (R,L)_\mu$
and separately under the non-spatial interchanges $\CP$: $\xi_{L,R} \leftra
\xi_{R,L}^\dagg$, $\xi_M \leftra \xi_M^\dagg$, $\Sigma_i \leftra
\Sigma_i^\dagg$, and $R_\mu^\alpha \leftra L_\mu^\alpha$. This $\CP$ can be
enlarged to include electromagnetic interactions: keeping $A_\mu = \sin\thw
W_\mu^3 + \cos\thw B_\mu$ while setting other electroweak gauge fields to
zero, $\Lsig$ remains invariant under $\CP$ with $eA_\mu \ra eA_\mu$. Thus,
this symmetry forbids, e.g., $\tpiz \ra \gamma\gamma$ and $\troz,\tom \ra
\gamma\tpiz$ and $\gamma Z^0$. As in QCD, there is no reason to expect that
TC respects this symmetry and these decays should occur. The WZW interaction
discussed in Sec.~IV violates $\CP$ and induces these processes.

The complete effective Lagrangian is
\be\label{eq:Leff}
\Leff = \Lsig + \LFF + \Lff + \LWZW + \CL_{M_\pi^2} + \Lpifbf\,.
\ee
The gauge-field Lagrangian has the standard form,
\be\label{eq:Lff}
\LFF = -{\tfourth} \left[W_{\mu\nu}^a W^{a,\,\mu\nu} + B_{\mu\nu} B^{\mu\nu}
  + R_{\mu\nu}^\alpha R^{\alpha,\,\mu\nu} + L_{\mu\nu}^\alpha
  L^{\alpha,\,\mu\nu}\right]\,,
\ee
where $a=1,2,3$. Quark and lepton couplings to gauge bosons involve {\em
  only} the primordial $W^a$ and $B$. This is important in controlling the
  energy dependence of $SM \ra SM$ amplitudes and in calculating the oblique
  parameters $S,T,W,Y$ in Sec.~VII. With an obvious condensed notation,
\be\label{eq:Lfbf} \Lff = \sum_{j=1}^3\Bigl[\bar
\psi_{jL}i\gamma^\mu D_{j\mu} \psi_{jL} + \bar u_{jR} i\gamma^\mu
D_{u_{jR}}u_{jR} + \bar d_{jR} i\gamma^\mu D_{d_{jR}}d_{jR} \Bigr]\,.
\ee
As shown for Higgsless models in Refs.~\cite{Chivukula:2005bn,
  Chivukula:2005xm,Chivukula:2005ji}, it is possible to reduce the value of
the $S$-parameter by introducing special couplings of the standard model
fermions to the $L$ and $R$ gauge bosons.  However, these couplings must be
finely tuned, and this is antithetical to our technicolor philosophy.
Finally, the Lagrangian $\CL_{M_\pi^2}$ includes ETC-induced $\tpi$ masses
and $\Lpifbf$ the $\tpi$ couplings to fermion-antifermion. They are discussed
in Sec.~V.\footnote{The shifts in $W^a$ and $B$ discussed in
  Eq.~(\ref{eq:Gshift}) also induce $\tpi$ couplings to quarks and leptons.
  These are also discussed in Sec.~V.}

\section*{III. Vector Boson States and Masses}

To transform to the unitary gauge, first make an $SU(2)_W$ transformation
with $U_W = \Sigma_2^\dagg(x)$, bringing $\Sigma_2$ to the identity, $\xi_L$
to $\xi_L' = \Sigma_2^\dagg \xi_L$, and $\xi_M$ and $\xi_R$ unchanged, so
that $\Sigma_1' = \Sigma_2^\dagg \Sigma_1$. Then make $U(2)_L$ and $U(2)_R$
transformations with $U_L = \xi_L'$ and $U_R = \xi_R^{'\,\dagg} \equiv
\xi_R^\dagg$. This takes those two fields to the identity and $\xi_M'$ and
$\Sigma_1'$ to
\bea\label{eq:newsigma}
\xi_M{''} &=& \xi_L' \xi_M' \xi_R' = \Sigma_1'\,;\\
\Sigma_1^{''} &=& \Sigma_1'\ 
  \equiv \exp{(2it\cdot\tilde\pi/F_1)}\,.
\eea
In the second equation, $\tilde\pi_\alpha$ are the
not-yet-canonically-normalized LSTC technipions. We relate them to
${\bs{\pi_T}}$ and $\tpipr$ in Eq.~(\ref{eq:canonical}) below. Dropping the
primes, $\Lsig$ now has the form
\bea\label{eq:Lsigtwo}
\Lsig = && {\tfourth} F_2^2\, \Tr\bigl|g\bs{t}\cdot\bs{W}_\mu - g' t_3
            B_\mu\bigr|^2 
      + {\tfourth} F_1^2 \Bigl\{a\Tr\bigl|i\partial_\mu\Sigma_1
              + g\bs{t}\cdot\bs{W}_\mu\Sigma_1 - 
              g' \Sigma_1 t_3 B_\mu\bigr|^2 \nn\\ 
      && \,+\, b\Bigl[\Tr\bigl|g\bs{t}\cdot\bs{W}_\mu + g'y_1 t_0 B_\mu
                   - g_T t\cdot L_\mu \bigr|^2 +\Tr\bigl|g'(t_3 + y_1 t_0)B_\mu
                - g_T t\cdot R_\mu\bigr|^2 \Bigr] \nn\\
        && \,+\, c\, \Tr\bigl|-i\partial_\mu\Sigma_1 + g_T(\Sigma_1 t\cdot R_\mu
        - t\cdot L_\mu \Sigma_1)\bigr|^2 \\ 
        && \,+\, d\, \Tr\bigl[(g\bs{t}\cdot\bs{W}_\mu \Sigma_1 - g'\Sigma_1
        t_3 B_\mu + g_T(\Sigma_1 t\cdot R_\mu - t\cdot L_\mu\Sigma_1) \nn\\
        &&\qquad\,\, \times (-i\partial_\mu\Sigma_1 + g_T(\Sigma_1 t\cdot
        R_\mu - t\cdot L_\mu\Sigma_1))\bigr]\nn\\
        && \,-\frac{if}{2g_T}\, \Tr\bigl[(-i\partial_\mu\Sigma_1 + 
            g_T(\Sigma_1 t\cdot R_\mu - t\cdot L_\mu  \Sigma_1))\Sigma_1^\dagg
                              (-i\partial_\nu\Sigma_1 + 
            g_T(\Sigma_1 t\cdot R_\nu - t\cdot L_\nu  \Sigma_1))\Sigma_1^\dagg
            t\cdot L^{\mu\nu} \nn\\  
        &&\qquad\,\, + \Sigma_1^\dagg(-i\partial_\mu\Sigma_1 + 
            g_T(\Sigma_1 t\cdot R_\mu - t\cdot L_\mu \Sigma_1))
                       \Sigma_1^\dagg(-i\partial_\nu\Sigma_1 + 
            g_T(\Sigma_1 t\cdot R_\nu - t\cdot L_\nu \Sigma_1))t\cdot R^{\mu\nu}
            \bigr] \Bigr\} \,.\nn
\eea

The charged and neutral gauge boson mass matrices can be read off from
$\Lsig$ by putting $\Sigma_1 \ra 1$. Defining
\be\label{eq:xsq}
x^2 = \frac{g^2}{2g_T^2}\,,
\ee
the charged mass matrix is (with rows and columns labeled, in order, by the
primordial $W^\pm = (W^1 \mp i W^2)/\sqrt{2}$, $V^\pm = (R^1 + L^1 \mp i (R^2
+ L^2))/2$, $A^\pm = (R^1 - L^1 \mp i (R^2 - L^2))/2$):
\be\label{eq:Mpm}
M_{\pm}^2 ={\tfourth} g_T^2 F_1^2
\left(\ba{ccc}
  2x^2\left(\frac{F_2^2}{F_1^2} + a + b\right)& -x b &
  x(b+d)\\ 
  -x b&  b&  0\\
x(b+d) &  0 & b+2(c+d)\ea\right) \ts.
\ee
The $5\times 5$ neutral mass matrix has rows and columns labeled by $W^3$,
$B$, $V^3 = (R^3+L^3)/\sqrt{2}$, $V^0 = (R^0+L^0)/\sqrt{2}$ and $A^3 =
(R^3-L^3)/\sqrt{2}$. The isoscalar axial vector $A^0 \equiv f_T$ does not mix
with these and, consequently, will not be produced as an $s$-channel
resonance in colliders. For this reason, we will not study its phenomenology
in this paper. The neutral mass matrix is
\begin{changemargin}{-1.0cm}{-.5cm}
   %
   %
\be\label{eq:Mzero}
M_0^2 ={\tfourth} g_T^2 F_1^2
 \left(\ba{ccccc}
  2x^2\left(\frac{F_2^2}{F_1^2} + a + b\right) &
 -2x^2\left(\frac{F_2^2}{F_1^2} + a\right)t_W & -bx & 0 & (b+d)x \\
 -2x^2\left(\frac{F_2^2}{F_1^2} + a\right)t_W & 
  2x^2\left(\frac{F_2^2}{F_1^2} + a + b(1+2y_1^2)\right)t_W^2 &
 -bxt_W &  -2bxy_1t_W & -(b+d)xt_W\\
 -bx &  -bxt_W &  b & 0 & 0 \\
  0 &  -2bxy_1t_W & 0 &  b  &  0\\
 (b+d)x & -(b+d)xt_W & 0 & 0 & b + 2(c+d)\ea\right) \,,
\ee
\end{changemargin}
where $t_W \equiv \tan\thw = g'/g$. This matrix has a zero-mass eigenstate,
the photon. The $f_T$ mass is
\be\label{eq:htmass}
M_{f_T}^2 = {\tfourth}g_T^2 (b + 2 (c+d)) F_1^2\,.
\ee
So long as $|a|,\dots,|d|$ are at most $\CO(1)$ and $F_2^2 \gg F_1^2$, we see
that $M_{V,A}^2 \sim {\fourth} g_T^2 F_1^2$ and $M_W^2 \sim {\fourth} g^2
F_2^2$. Then, in order that $M_{V,A}^2 \gg M_{W,Z}^2$
\be\label{eq:magnitudes}
\frac{g_T^2}{g^2} \gg \frac{F_2^2}{F_1^2} \gg 1 \gg x^2\,.
\ee
From $M_{\pm,0}^2$ we can read off the approximate mixings of the
technivectors with the primordial electroweak bosons --- $W^\pm$, photon $A =
W^3\sin\thw + B\cos\thw$, and $Z = W^3\cos\thw - B\sin\thw$:\footnote{All the
  gauge bosons are canonically normalized.}
\bea\label{eq:approxmixings}
&& f_{AV^3} \simeq \frac{M^2_{A V_3}}{M^2_{V_3 V_3}} =
-2x\sin\thw = -\frac{\sqrt{2}e}{g_T}\,,\quad
f_{AV^0} \simeq  -\frac{\sqrt{2}ey_1}{g_T}\,,\quad
f_{AA^3} = 0\,;\\
&& f_{ZV^3} \simeq -\frac{\sqrt{2}e\cot 2\thw}{g_T}\,,\,\,
f_{ZV^0} \simeq \frac{\sqrt{2}ey_1\tan\thw}{g_T}\,,\,\,
f_{ZA^3} \simeq \frac{\sqrt{2}eD}{Bg_T\sin 2\thw}\,;\\
&& f_{W^\pm V^\mp} \simeq -\frac{e}{\sqrt{2}g_T\sin\thw}\,,\quad
f_{W^\pm A^\mp} \simeq \frac{eD}{\sqrt{2}Bg_T\sin\thw}\,;
\eea
where $e = g\sin\thw = g'\cos\thw$ and $y_1 = Q_U + Q_D$. For convenience, we
are introducing the following combinations of $a,b,c,d$:
\bea\label{eq:etaGfactors}
A &=& a(b+2(c+d)) + bc -\thalf d^2 \equiv aB + bc - \thalf d^2\,,\nn\\
B &=& b + 2(c+d)\,, \qquad C = 2c+d\,,\qquad D = b+d \equiv B - C\,.
\eea
The mixing parameters in the TCSM corresponding to
Eqs.~(\ref{eq:approxmixings}) are (from Refs.~\cite{Lane:2002sm,
  Eichten:2007sx}):\footnote{The signs of these $f$'s are opposite those in
  these references; their overall sign is purely a matter of convention.}
$f_{\gamma\troz} = -e/g_{\tro}$, $f_{\gamma\tom} = -ey_1/g_{\tro}$,
$f_{Z\taz} = e/g_{\ta}\sin 2\thw$, etc.
The coupling of $\troz$ to the weak isospin current, $j_\mu^3$, is
$M_{\tro}^2/g_{\tro}$ and the coupling of $\taz$ to $j_{5\mu}^3$ is
$M_{\ta}^2/g_{\ta}$. Then, to leading order in $x$, $\Leff$ produces the TCSM
mixings if we identify the HLS gauge coupling $g_T$ to be
\be\label{eq:grelations}
g_T \simeq \sqrt{2}g_{\tro}\simeq \frac{\sqrt{2}D}{B}g_{\ta}\,.
\ee
One numerical estimate of $g_{\tro}$ may be obtained, rather cavalierly, by
scaling from QCD using large-$\Ntc$. Using the QCD value $\alpha_\rho =
2.16$, extracted from the rate for $\tau \ra \rho
\nu_\tau$~\cite{Amsler:2008zzb}, this gives $g_{\tro} =
\sqrt{4\pi(2.16)(3/\Ntc)}$. With this identification,
\be\label{eq:xsqest}
x^2 \cong \alpha_{EM}/(4\atro\sin^2\thw) = 0.52 \times 10^{-2}
\ee
for $N_{TC} = 4$.

The condition $g_{\ta} \cong g_{\tro}$ that the $F_1$-scale contribution
$S_1$ to the $S$-parameter in Eq.~(\ref{eq:Snarrow}) is $B\cong D$, i.e.,
\be\label{eq:Scondition}
C = 2c + d \cong 0
\ee
We shall confirm this in Sec.~VII. The condition $M_{\ta} \cong M_{\tro}$
(which, strictly speaking, we don't need for small $S_1$) implies that $c+d
\cong 0$. Together, these are the condition $c=d=0$ used in the DBESS model
to make the $S$-parameter small~\cite{Casalbuoni:1995qt}. The enhanced
symmetry implied by this condition is discussed in
Ref.~\cite{Appelquist:1999dq}.  Neither of these papers employed the
$f$-interaction, so their conclusions about the consequences of $c=d=0$ for
$\tro \ra WW,\,WZ$ do not apply to us.

Diagonalizing the charged mass matrix through $\CO(x^2)$ and for $c,d \neq
0$, we obtain
\bea\label{eq:mcharev}
M_{W^\pm}^2 &=& {\tfourth} g^2\biggl[F_2^2 +\frac{A}{B}F_1^2 \biggr] \equiv
{\tfourth} g^2  F_\pi^2\,,\nn\\
M_{\tropm}^2 &=& {\tfourth} g_T^2 F_1^2 b(1 + x^2)\,,\nn\\
M_{\tapm}^2 &=& {\tfourth} g_T^2 F_1^2B\left(1 + \frac{x^2D^2}{B^2}\right)\,,
\eea
where we introduced the fundamental electroweak scale of the LSTC described
by $\Leff$, $F_\pi = \sqrt{F_2^2 + AF_1^2/B} = 246\,\gev$. The ``mixing
angle'' $\chi$ characterizing the contribution of the low $F_1$-scale to
electroweak symmetry breaking is
\be\label{eq:sinchi}
\sin\chi = \sqrt{\frac{A}{B}}\frac{F_1}{F_\pi}\,.
\ee
Note the additional factor of $\sqrt{A/B}$ (expected to be $\CO(1)$) relative
to the TCSM definition, $F_1/F_\pi$. This is due to our having defined $F_1$
as the decay constant of the non-canonically normalized $\tilde \pi$-fields
in Eq.~(\ref{eq:newsigma}) (see Eq.~(\ref{eq:canonical}) below). The nonzero
neutral eigenmasses, through $\CO(x^2)$ and $\CO(y_1^2\sin^4\thw)$, are given
by (again, for $c,d \neq 0$)
\bea\label{eq:mneutev}
M_{Z^0}^2 &=& {\tfourth} (g^2+g^{'2})F_\pi^2 =
\frac{M_{W^\pm}^2}{\cos^2\thw}\,,\nn\\
M_{\troz}^2 &=& {\tfourth} g_T^2 F_1^2 b\biggl[1 +
\frac{x^2(1+4y_1^2\sin^4\thw)}{\cos^2\thw}\biggr]\,, \nn\\
M_{\tom}^2 &=&  {\tfourth} g_T^2 F_1^2 b(1 + 4x^2 y_1^2
\sin^2\thw)\,, \nn\\
M_{\taz}^2 &=&  {\tfourth} g_T^2 F_1^2 B\biggl[1 +
\biggl(\frac{xD}{B\cos\thw}\biggr)^2\biggr]\,, \nn\\
M_{f_T}^2 &=&   {\tfourth} g_T^2 F_1^2 B \,.
\eea
Note that the zeroth-order $V_3 \cong \troz$ and $V_0 \cong \tom$ masses are
equal, $\half g_T \sqrt{b} F_1$ and are split only by terms of $\CO(x^2)$.
Thus, the phenomenology of our $\Leff$ has very nearly degenerate $\tom$ and
$\troz$. If we wish to split them by more than $\CO(x^2)$, it is necessary to
use $U(1)_{L,R}$ couplings $g_T' \neq g_T$. That is an easy modification to
adopt, but we shall not do so in this paper. The eigenvectors in the charged
and neutral sectors are in Appendix~B.\footnote{The point $c=d=0$ is a
  singular one for the mass matrices. In that case, the charged eigenstates
  are slightly ($\CO(x)$) mixed $W^\pm$ and $L^\pm$ with masses $M_W^2 =
  \tfourth g^2 F_\pi^2$ and $M_{L^\pm}^2 \cong \tfourth g_T^2 F_1^2
  b(1+2x^2)$, and $R^\pm$ with mass $M_{R^\pm}^2 = \tfourth g_T^2 b$. The
  neutral eigenstates are the massless photon, slightly mixed $Z$ and $L^3$
  with masses $\tfourth g^2 F_\pi^2/\cos^2\thw$, $R^3$ with mass $\tfourth
  g_T^2 F_1 b(1+2x^2)$ and a degenerate $\tom$ and $f_T$ with mass
  $M_{\tom,f_T}^2 = \tfourth g_T^2 F_1^2 b$.}

The last step in preparing the Lagrangian with properly normalized fields
requires eliminating gauge-technipion kinetic terms. In going to unitary
gauge, we removed mixing between gauge and unphysical Goldstone bosons, but
not those involving the $\tpi$. To eliminate $G_\mu\partial^\mu\tpi$ terms,
we shift the gauge fields by linear functions $\zeta$ of the $\tilde \pi_T$:
\be\label{eq:Gshift}
g_G G^\alpha_\mu \ra g_G G^\alpha_\mu + \partial_\mu \zeta_{G^\alpha}/F_1\,,
\quad (G^\alpha = W^a,B,V^\alpha =
(R+L)^\alpha/\sqrt{2},A^\alpha=(R-L)^\alpha/\sqrt{2})\,,
\ee
Unlike the transformation to unitary gauge, the Lagrangian is not invariant
under these shifts. Therefore, we must include them in all the terms in
Eq.~(\ref{eq:Leff}). Once the shifts are done, we can read off the
coefficients of ${\half} (\partial_\mu\tpi)^2$ and scale the $\tpi$
appropriately. The shift fields $\zeta_{G^\alpha}$ are in Appendix~B. The
$\tilde\pi_\alpha$ are related to the canonically-normalized $\pi_\alpha
\equiv ({\bs{\pi_T}},\tpipr)$ by
\be\label{eq:canonical}
\tilde\pi_a = \frac{\sqrt{B/A}}{\cos\chi}\pi_a\ \equiv \eta\pi_a,, \quad
\tilde\pi_0 = \sqrt{\frac{B}{A}}\pi_0\
\equiv \eta'\pi_0,.
\ee

Finally, we record the electroweak parameters $e_R^2$ and $(\tan^2\thw)_R$
through $\CO(x^2)$, are
\bea\label{eq:renorm}
\frac{1}{e_R^2} &=& \frac{1}{e^2}[1 + 4x^2(1+y_1^2)\sin^2\thw]\nn\\
(\tan^2\thw)_R &=& \tan^2\thw\left[1 + x^2\left(1  + 2\cos 2\thw - 4 y_1^2
    \sin^2\thw - \frac{D^2}{B^2}\right)\right]\,.
\eea

\section*{IV. The Wess-Zumino-Witten Interaction}

As we discussed in Sec.~II, the HLS interaction $\Lsig$ has a symmetry,
$\CP$, that forbids $\tpiz \ra \gamma\gamma$ and $\tom,\,\tro \ra \gamma
\tpi$. The interaction's $SU(2)$ gauge structures also forbid $\tom \ra
\gamma Z^0$ and $\tro \ra \gamma Z^0,\, \gamma W$. There is no reason not to
expect such decays in LSTC and, moreover, they may be of considerable
phenomenological importance. For example, $\tom \ra \gamma Z^0$ is likely to
be the discovery channel for $\tom$ at the LHC~\cite{Brooijmans:2008se}. Such
processes might be found in $\CP$-violating Wess-Zumino-Witten interactions
induced by anomalously-nonconserved symmetries of the underlying TC theory.

It is, in fact, not clear how to construct $\LWZW$ for the theory whose
chiral Lagrangian is $\Lsig$. A general approach for discussing the WZW terms
for an effective theory of pions and vector and axial-vector mesons was
developed by Kaymackcalan, Rajeev and Schechter~\cite{Kaymakcalan:1983qq} and
by Harvey, Hill and Hill (HHH)~\cite{Harvey:2007ca}.
Ref.~\cite{Kaymakcalan:1983qq} was concerned with the electromagnetic
interactions of these mesons; HHH generalized this to include full $(SU(2)
\otimes U(1))_{EW}$ gauge invariance. The situation studied by HHH is similar
to ours. They considered the standard model with one doublet each of quarks
and leptons, and addressed the question of constructing $\LWZW$ when the
quarks had been integrated out. Their effective Lagrangian describes the
$U(2)_L \otimes U(2)_R$-invariant interactions of pions and
$\rho,\omega,a_1,f$. They treated the spin-one mesons as {\em nondynamical}
background fields. The essence of Ref.~\cite{Harvey:2007ca} is the
determination of counterterms needed to maintain the local $SU(2)\otimes
U(1)$ invariance of $\LWZW$ in the presence of the global $U(2)_L \otimes
U(2)_R$ symmetry. If we follow their method exactly, our WZW action would be
given by their Eqs.~(69) with $\CA_{L,R} = A_{L,R} + B_{L,R}$ where $A_{L,R}$
are the appropriate $SU(2)\otimes U(1)$ fields, $W$ and $B$, and $B_{L,R}$
are the $U(2)_{L,R}$ background fields $L$ and $R$. One important difference
is that their $\Gamma_{AAA}$ and $\Gamma_{AAAA}$ would be absent from our WZW
action because we integrated out {\em all} the technifermions and so there
are no anomalies associated with the electroweak symmetries.

However, this approach is inappropriate for us. Our $L$ and $R$ are dynamical
fields, not backgrounds. More importantly, if we think of $L$ and $R$ as
composites of underlying fermions, the ones composing $\xi_{L,M,R}$, those
fermions are not just our just light technifermion doublet,
$(T_U,T_D)_{L,R}$, whose isospin indices are gauged in the electroweak
group and not in $U(2)_{L,R}$. Furthermore, the $U(2)_L\otimes U(2)_R$ gauge
currents composed of the additional fermions are anomalously nonconserved.
There is nothing to cancel these anomalies, and the HLS gauge interaction is
nonrenormalizable. To our knowledge, the problem of determining $\LWZW$ for
such a theory has not been discussed before.

A second approach we considered, therefore, follows HHH, but $\LWZW$ was
constructed just for the $U(2)_L\otimes U(2)_R$ center of the moose in
Fig.~\ref{moose}, i.e., treating $L$ and $R$ as dynamical ($A$-type), not
background ($B$-type) gauge fields. The motivation for this is that, having
integrated out $(T_U,T_D)$, the only anomalies of the theory should be those
associated with $U(2)_L \otimes U(2)_R$ and that of the baryon number
current, $\ol T_U \gamma_\mu T_U + \ol T_D \gamma_\mu T_D$. This procedure
fails because it breaks electroweak gauge invariance. Since $L,R$ are
dynamical fields mixing with the $SU(2)\otimes U(1)$ fields, the
mass-eigenstate electroweak bosons, including the photon, are admixtures
involving $R \pm L$. The breakdown of EM gauge invariance is manifest in this
$\LWZW$; it contains terms inducing $\taz,\, f_T \ra \gamma \gamma$, a
violation of Yang's theorem.

To circumvent that problem, we employed HHH's procedure on each of the three
sub-mooses in Fig.~\ref{moose}. That is, we used their Eq.~(69), successively
taking $\CA_{L,R} = A_{L,R}$ with $A_{L\mu} = W_\mu$ and $A_{R\mu} = L_\mu$;
$A_{L\mu} = L_\mu$ and $A_{R\mu} = R_\mu$; $A_{L\mu} = R_\mu$ and $A_{R\mu} =
B_\mu$ {\em including}, implicitly, the shifts as in
Eq.~(\ref{eq:Gshift}).\footnote{We did not close the moose in a circle by
  also taking $A_{L\mu} = W_\mu$ and $A_{R\mu} = B_\mu$ because, having
  integrated out all the technifermions coupling to these fields, the
  corresponding $\Gamma_{AAA} = \Gamma_{AAAA} = 0$.} We then added the
resulting WZW interactions. This calculation was done in the unitary gauge in
which $\xi_M = \Sigma_1 = \exp{(2it\cdot\tilde\pi/F_1)}$.

Several phenomenologically important WZW interactions resulting from our
procedure are presented in Sec.~VI. Here we discuss the interaction for the
isovector $\tpiz \ra \gamma \gamma$. Its strength determines that of all the
other WZW interactions. It is given by
\be\label{eq:pitogg}
\CL(\tpiz \ra \gamma \gamma) = -\frac{e^2 y_1\Ntc (1 -
  {\textstyle{\frac{2}{3}}}\sin^2\chi)}{16\pi^2 \sqrt{A/B}\,F_1\cos\chi}\, \tpiz 
F_{\mu\nu} \widetilde F_{\mu\nu}
\,,
\ee
where $\widetilde F_{\mu\nu} = \half \epsilon_{\mu\nu\rho\sigma}
F^{\rho\sigma}$ and we assumed that the fermions $(T_U,T_D)$ transform
according the fundamental ${\bs N}_{TC}$ representation of $\sutc$. This is
just what we expect to leading order in $\sin^2\chi$. In QCD, the coefficient
of $\pi^0 F_{\mu\nu} \widetilde F_{\mu\nu}$ is $\thalf \cdot N_C \cdot (Q_u^2
- Q_d^2) e^2/(8\pi f_\pi)$. The corresponding factor here is $\thalf \cdot
\Ntc \cdot y_1 e^2/(8\pi\sqrt{A/B} F_1 \cos\chi)$ where we used
Eq.~(\ref{eq:canonical}), $\pi_a = \sqrt{A/B}\cos\chi\,\tilde\pi_a$, and
$\langle\Omega|\thalf \ol T\gamma_\mu\gamma_5 \tau_3 T|\tilde\pi_3(q)\rangle
= i F_1 q_\mu$.

\section*{V. Technipion Masses and Couplings to Fermions}

Technipion masses are generated mainly by ETC
interactions~\cite{Eichten:1979ah}. As in the TCSM, we assume for simplicity
that technipion masses are isospin symmetric but, as explained earlier, there
is no need for $M_{\tpipr}$ to equal $M_{\tpi}$. We describe their masses by
the simple Lagrangian
\be\label{eq:Mtpi}
\CL_{M_\pi^2} = -{\frac{1}{4\eta^2}} M_1^2 F_1^2 \Tr(\Sigma_1 +
    \Sigma_1^\dagg) + {\frac{1}{32\eta^{'2}}} M_2^2 F_1^2
    |\Tr(\Sigma_1 - \Sigma_1^\dagg)|^2\,,
\ee
where $\eta$ and $\eta'$ are the normalization constants of
Eq.~(\ref{eq:canonical}). Then,
\be\label{eq:pimasses}
M_{\tpi}^2 = M_1^2\,\qquad M_{\tpipr}^2 = M_1^2 + M_2^2\,.
\ee
We shall assume that $M_2^2 \gg M_1^2$ and not discuss $\tpipr$ phenomenology
further.

Technipion decays to fermion pairs are also induced by ETC interactions. In
the absence of an explicit ETC model, we can only guess at the form of the
$\tpi$-decay Lagrangian. Because the same ETC bosons induce $\tpi \bar f_{iL}
f_{jR}$ and the $\bar f_{iL} f_{jR}$ mass term, we {\em expect} that the
couplings are Higgslike, i.e., approximately proportional to the fermions'
masses. To maintain consistency with the way the decays are modeled in {\sc
  Pythia}, we take the effective Lagrangian for the coupling of a technipion
to a pair of fermions $f_i \bar f_j$ with masses $m_i$, $m_j$ (renormalized
at the mass of the technipion) to be
\be\label{eq:Lpiff}
\Lpifbf = \sum_{\tpi,i,j}
\frac{C_{\tpi,ij}(m_i+m_j)}{\sqrt{A/B}F_1\cos\chi}\, \tpi \bar f_{iL} f_{jR} +
{\rm h.c.}\,.
\ee
Here, we assume that $C_{\tpi,ij}$ is a constant of $\CO(1)$, {\em without}
CKM-like mixing angle suppression, {\em unless} one or both
fermions are top quarks. For light fermions, $C_{\tpi,ij} = 1$ if $\tpi =
\tpiz$ and $C_{\tpi,ij} = \sqrt{2}$ if $\tpi = \tpipm$. If either
fermion (or both) is a top quark, $m_t$ is to be replaced by $m_b$,
reflecting the fact that ETC interactions probably contribute at most
$\sim 5\,\gev$ to the top's mass~\cite{Hill:1994hp}.

The shifts of the primordial $W$ and $B$-fields shifts discussed in Sec.~III
induce another coupling of $\tpi$ to quarks and leptons. As can be seen from
Eq.~(\ref{eq:etaG}), they are of order $(m_i+m_j)\sin^2\chi(\sqrt{A/2B}
F_1\cos\chi)$ for the $W^\pm$ shift. Since $\sin^2\chi \ll 1$ in LSTC, they
can be important only for the $\tpi$ couplings to $t \bar b$ and $t \bar t$.
The CDF limit on $t \ra H^+ b$ (with $H^+$ assumed to decay to $c \bar s$) is
$B(t \ra H^+ b) \simle 0.20 \pm 0.10$~\cite{Collaboration:2009ke}. This puts
no meaningful restriction on $\sin\chi$ for the current CDF limit of
$M_{\tpipm} \simge 125\,\gev$~\cite{Nagai:2008xq}.

\section*{VI. $\Leff$ at the Tevatron and LHC and Comparison with the TCSM}

Walking technicolor dynamics probably close off the two and three-$\tpi$
decay channels of $\tro$, $\tom$ and $\ta$~\cite{Lane:1989ej}. This makes
them very narrow, with striking decay signatures and favorable signal to
background ratios. To repeat, we assume in this paper that the isosinglet
$\tpipr$ is too heavy to appear in technivector decays. Then, at the Tevatron
the most promising decay channels likely are $\tro \ra W \tpi$, $\tom \ra
\gamma \tpiz$, $\taz \ra W\tpi$ and and $\tapm \ra \gamma \tpi,\,W
\tpi,\,Z\tpi$~\cite{Lane:2002sm, Eichten:2007sx}. The weak bosons are sought
in their decay to electrons or muons and missing energy. Technipions
accessible at the Tevatron must be lighter than top quarks, so the expected
signatures there are $\tpipm \ra q \bar b$ and $\tpiz \ra b \bar b$. As
discussed in Sec.~V, the Tevatron limits on $t \ra H^+ b$ are consistent with
our assumption that the $\tpi$ coupling to the $t$-quark is small, probably
of order $m_b/F_1$.

At the LHC, the backgrounds to technivector decays to $\tpi$ plus an
electroweak gauge boson depend on how important the decay modes $\tpi \ra t
\bar q$ and $t \bar t$ are. If they are unimportant, hadronic backgrounds,
especially $\bar tt$ production, make $\tro \ra W \tpi \ra$ leptons $+$
$(b+q)$--jets unobservable.\footnote{At high luminosity, $\CO(100\,\ifb)$,
  $\trop \ra \tpip Z \ra q \bar b \ellp\ellm$ appears to be observable above
  background~\cite{Brooijmans:2008se}.} No studies have been carried out yet
on backgrounds when $\tpi \ra t \bar q$ is substantial. Thus, at the LHC, the
most promising discovery channels appear to be the low-rate, but relatively
background-free, modes $\tropm \ra W^\pm Z^0$, $\tom \ra \gamma Z^0$ and
$\tapm \ra \gamma W^\pm$, with $W,Z \ra e,\mu$--leptons. The weak bosons in
these decays are expected to be mainly longitudinally polarized, providing
technivector decay angular distributions that are indicative of their
underlying dynamical origin~\cite{Eichten:2007sx,Brooijmans:2008se}.

For use in calculating the important technivector decay rates, we record in
this section the relevant {\em on-mass-shell} operators from
$\Leff$.\footnote{The approximate forms given for these on-shell operators
  cannot be used to compute $s$-dependent widths nor as pieces of a
  scattering amplitude for, e.g., $u\bar d \ra W^+,\trop,\tap \ra W^+ Z^0$.}
They were calculated to leading order in $x^2 = g^2/2g_T^2$,
$\sin^2\chi = AF_1^2/BF_\pi^2$ and $y_1\sin^2\theta_W$. All fields are
mass-eigenstates. In general, we simplified the interactions by using
leading-order equations of motion such as $\partial_\mu \rho_{T\mu\nu} =
-M^2_{\tro}\rho_{T\nu} = -\tfourth bg_T^2F_1^2\rho_{T\nu}(1+\CO(x^2)) =
-\tfourth bg_T^2(B/A)\,F_\pi^2\sin^2\chi\,\rho_{T\nu}(1+\CO(x^2))$, where
$\rho_{T\mu\nu} = \partial_\mu\rho_{T\nu} - \partial_\nu\rho_{T\mu}$, and
$\partial_\mu \rho_{T\mu} = 0$, and by dropping total derivatives.

At the end of this section we will compare these decay operators with what is
expected from the TCSM. As we have said, it is not clear --- especially from
our consideration of the underlying theory's anomalies and the WZW
interactions they imply --- that our effective Lagrangian is based just
on a theory with a single technifermion doublet, $(T_U,T_D)$. Whether it is
or not (and the WZW discussion suggests it isn't), we find complete agreement
between these operators and those that occur in the TCSM summarized in
App.~A~\cite{Lane:2002sm, Eichten:2007sx}.

We start with the operators for $\troz$ two-body decays to technipions and
weak bosons.
\bea\label{eq:rzpipi}
\CL(\troz \ra \tpip\tpim) &=& -\frac{ig_T bCY}{2\sqrt{2}\,A}
\,\rho_{T\mu}^0 \, \tpip \olra{\partial_\mu} \tpim \nn\\
&& + \frac{2i[C(C\cos^2\chi + 2B\sin^2\chi) -fD^2\cos^2\chi]Y}
{\sqrt{2}\,g_T(B F_\pi\sin\chi)^2} 
\,\rho_{T\mu\nu}^0 \, \partial_\mu\tpip\,\partial_\nu\tpim \nn\\
&\cong& -i\grpp\cos^2\chi\,\rho_{T\mu}^0 \,
\tpip \olra{\partial_\mu} \tpim \,.
\eea
Here, $Y = (1-2y_1^2\sin^4\thw)$, the difference from unity being a measure
of weak isospin violation. Under the reasonable presumption that $Y \cong
1$, we defined the $\tro\tpi\tpi$-coupling
\be\label{eq:rpp}
\grpp = \frac{b[C(B+D) + fD^2]g_T}{4\sqrt{2}\,AB} \cong
\frac{M_{\tro}^2}{\sqrt{2}g_T F_\pi^2 \sin^2\chi}
\left[1 + (f-1)\frac{M_{A_2}^2}{M_{A_1}^2}\right]
\,,
\ee
where the mass parameters $M_{A_1}$ and $M_{A_2}$ are defined in
Eqs.~(\ref{eq:FFscaleone},\ref{eq:FFscaletwo}) below.
This corresponds to $g_{\tro}$ in Eq.~(7) of the TCSM
paper~\cite{Lane:2002sm} (and {\em not} necessarily to the coupling of $\tpi$
to the axial isospin current in Eq.~(\ref{eq:Snarrow})). In the above
approximations, the $\troz \ra W^\pm \tpimp$ and $W^+ W^-$ interactions are
(with $Y \cong 1$):
\bea\label{eq:rzWpi}
\CL(\troz \ra W^\pm\tpimp) &\cong&
\frac{igg_T bCF_\pi\sin\chi}{4\sqrt{2}\,A\cos\chi}\, \rho^0_{T\mu}(\tpip
W_\mu^- - \tpim W_\mu^+) \nn\\ 
&&+ \frac{ig CD\cos\chi}{\sqrt{2}\,g_T B^2 F_\pi\sin\chi}
[\rho_{T\mu\nu}^0(\partial_\mu \tpip W_\nu^- - \partial_\mu
\tpim W_\nu^+) +\rho_{T\nu}^0(W_{\mu\nu}^+ \partial_\mu \tpim - W_{\mu\nu}^-
\partial_\mu\tpip)] \nn\\
&& + \frac{ig fD^2\cos\chi}{\sqrt{2}\,g_T B^2 F_\pi\sin\chi}\,\rho_{T\mu\nu}^0
(\partial_\mu \tpip W_\nu^- - \partial_\mu \tpim W_\nu^+) \nn\\
&& - \frac{eg_T(B+D)\Ntc y_1 \sin\thw \cos\chi}{32\sqrt{2AB} \,\pi^2 F_1}
  \rho_{T\mu\nu}^0 (\widetilde W_{\mu\nu}^+ \tpim +  
\widetilde W_{\mu\nu}^- \tpip) \nn\\
&\cong& ig\grpp  F_\pi \sin\chi \cos\chi \, \rho_{T\mu}^0(\tpip
W_\mu^- - \tpim W_\mu^+)\nn \\
&& + \frac{ie f D^2  \cos\chi}{2\sqrt{2AB} B g_T F_1 \sin\thw} 
\rho_{T\mu\nu}^0  (W_{\mu\nu}^+ \tpim - W_{\mu\nu}^- \tpip)\nn\\
&& - \frac{eg_T(B+D)\Ntc y_1 \sin\thw \cos\chi}{32\sqrt{2AB} \,\pi^2 F_1}
  \rho_{T\mu\nu}^0 (\widetilde W_{\mu\nu}^+ \tpim +  
\widetilde W_{\mu\nu}^- \tpip)
\nn\\
&\longra& +i\grpp\sin\chi\cos\chi \,\rho_{T\mu}^0(\tpip
\olra{\partial_\mu}\Pi_T^- + \Pi_T^+ \olra{\partial_\mu}\tpim)\,;
\eea
where $\widetilde W_{\mu\nu} = \half \epsilon_{\mu\nu\rho\sigma}
W^{\rho\sigma}$ and, in accord with Eq.~(\ref{eq:magnitudes}), we dropped
terms of relative $\CO(M_W^2/M_{\tro}^2)$. Next,
\bea\label{eq:rzWW}
\CL(\troz \ra W^+W^-) &\cong& -\frac{ig^2 C(B+D)}{2\sqrt{2}\,g_T
  B^2}[\rho_{T\mu\nu}^0 W_\mu^+ W_\nu^- + \rho_{T\mu}^0(W_\nu^+ W_{\mu\nu}^-
- W_\nu^- W_{\mu\nu}^+)]\nn\\
&& - \frac{ig^2 fD^2}{2\sqrt{2}\,g_T B^2}\,\rho_{T\mu\nu}^0 W_\mu^+ W_\nu^-
\nn\\ 
&& + \frac{egg_T(B+D) \Ntc y_1 \sin\thw}{32\sqrt{2} B \,\pi^2} \rho_{T\mu}^0
(W_\nu^- \widetilde W_{\mu\nu}^+ + W_\nu^+ \widetilde W_{\mu\nu}^-) \nn\\ 
&\longra& -i\grpp\sin^2\chi\,\rho_{T\mu}^0
\,\Pi_T^+\olra{\partial_\mu}\Pi_T^-  \nn\\
&& - \frac{ie f D^2  \cos\chi}{2\sqrt{2AB} B g_T F_1 \sin\thw} 
\rho_{T\mu\nu}^0  (W_{\mu\nu}^+ \Pi_T^- - W_{\mu\nu}^- \Pi_T^+)\nn\\
&& + \frac{eg_T(B+D)\Ntc y_1 \sin\thw \sin\chi}{32\sqrt{2AB} \,\pi^2 F_1}
  \rho_{T\mu\nu}^0 (\widetilde W_{\mu\nu}^+ \Pi_T^- +  
\widetilde W_{\mu\nu}^- \Pi_T^+)\,;
\eea
\bea\label{eq:rzZpi}
\CL(\troz \ra Z^0 \tpiz) &\cong& 
\frac{eg_T(B+D) \Ntc y_1 \sin^2\thw \tan\thw \cos\chi}
{16\sqrt{2AB} \,\pi^2 F_1}\, \rho_{T\mu\nu} \widetilde Z_{\mu\nu}^0
\tpiz \,;
\eea
\bea\label{eq:rzZZ}
\CL(\troz \ra Z^0 Z^0) &\cong& 
-\frac{g^2 g_T(B+D) \Ntc y_1 \sin^2\thw \tan^2\thw}{16\sqrt{2}B \,\pi^2}\,
\rho_{T\mu}^0 Z_\nu^0 \widetilde Z_{\mu\nu}^0\nn\\
&\longra& -\frac{eg_T(B+D) \Ntc y_1 \sin^2\thw \tan\thw \sin\chi}
{16\sqrt{2AB} \,\pi^2 F_1}\, \rho_{T\mu\nu} \widetilde Z_{\mu\nu}^0
\Pi_T^0 \,.
\eea
Note the $f$-term $\rho_{T\mu\nu} W_{\mu\nu} \tpi$ and WZW $\rho_{T\mu\nu}
\widetilde W_{\mu\nu} \tpi$ forms in Eqs.~(\ref{eq:rzWpi}--\ref{eq:rzZZ}).
These correspond to the $FF\tpi$ and $F \widetilde F\tpi$ terms in
Eq.~(\ref{eq:amplitudes}) of App.~A and will be discussed below. In
Eqs.~(\ref{eq:rzWpi},\ref{eq:rzWW},\ref{eq:rzZZ}) we indicated their ``TCSM
limit''. In that limit, amplitudes involving weak gauge bosons are dominated
at large $M_{\tro}/M_W$ by the emission of their longitudinally-polarized
components with $W_{L\mu}^{\pm,0} \cong \partial_\mu \Pi_T^{\pm,0}/M_W =
2\partial_\mu\Pi_T^{\pm,0}/(gF_\pi)$, where $\Pi_T$ is the unphysical
Goldstone boson. Note that $W_{\mu\nu}$ has no large $W_L$-piece.

The corresponding charged $\tro$ decay operators are:
\bea\label{eq:rpmpipi}
\CL(\tropm \ra \tpipm\tpiz) &\cong& -\frac{ig_T bC}{2\sqrt{2}\,A}
[\rho_{T\mu}^+ \, \tpim \olra{\partial_\mu} \tpiz + \rho_{T\mu}^- \, \tpiz
\olra{\partial_\mu} \tpip]\nn\\
&& + \frac{2i[C(C\cos^2\chi + 2B\sin^2\chi) -fD^2\cos^2\chi]}
{\sqrt{2}\,g_T(B F_\pi\sin\chi)^2} 
[\rho_{T\mu\nu}^+ \, \partial_\mu\tpim\,\partial_\nu\tpiz  + \rho_{T\mu\nu}^-
\,\partial_\mu\tpiz\,\partial_\nu\tpip]\nn\\
&\cong& -i\grpp\cos^2\chi[\rho_{T\mu}^+ \,\tpim \olra{\partial_\mu} \tpiz +
\rho_{T\mu}^- \,\tpiz \olra{\partial_\mu} \tpip]\,.
\eea
and
\bea\label{eq:rpmWpi}
\CL(\tropm \ra W^\pm\tpiz) &\cong&
-ig\grpp F_\pi \sin\chi \cos\chi (\rho_{T\mu}^+ W_\mu^- -
\rho_{T\mu}^- W_\mu^+)\tpiz \nn\\
&& + \frac{ie f D^2 \cos\chi}{2\sqrt{2AB} B g_T F_1 \sin\thw} 
(\rho_{T\mu\nu}^+ W_{\mu\nu}^- - \rho_{T\mu\nu}^- W_{\mu\nu}^+)\tpiz \nn\\
&\longra& i\grpp\sin\chi\cos\chi(\rho_{T\mu}^+\Pi_T^- \olra{\partial_\mu}\tpiz +
\rho_{T\mu}^-\tpiz \olra{\partial_\mu}\Pi_T^+)\,;
\eea
and
\bea\label{eq:rpmZpi}
\CL(\tropm \ra Z^0\tpipm) &\cong&
ig\grpp F_\pi \sin\chi \cos\chi (\rho_{T\mu}^+ \tpim -
\rho_{T\mu}^- \tpip)Z_\mu^0 \nn\\
&& - \frac{ief D^2 \cos\chi}{\sqrt{2AB} B g_T F_1 \sin 2\thw} 
(\rho_{T\mu\nu}^+ \tpim - \rho_{T\mu\nu}^- \tpip)Z_{\mu\nu}^0 \nn\\
&& + \frac{eg_T(B+D)\Ntc y_1 \tan\thw \cos\chi}{32\sqrt{2AB} \,\pi^2 F_1}
(\rho_{T\mu\nu}^+ \tpim + \rho_{T\mu\nu}^- \tpip)\widetilde Z_{\mu\nu}^0 \nn\\ 
&\longra& i\grpp\sin\chi\cos\chi(\rho_{T\mu}^+\tpim \olra{\partial_\mu}\Pi_T^0 +
\rho_{T\mu}^-\Pi_T^0 \olra{\partial_\mu}\tpip)\,;
\eea
and
\bea\label{eq:rpmWZ}
\CL(\tropm \ra W^\pm Z^0) &\cong& -\frac{ig^2 C(B+D)}{2\sqrt{2}\,g_T B^2\cos\thw}
[(W_\mu^+ \rho_{T\nu}^- -W_\mu^- \rho_{T\nu}^+)Z^0_{\mu\nu} \nn\\
&&+ (\rho_{T\mu\nu}^+ W_\mu^- - \rho_{T\mu\nu}^- W_\mu^+ +
     W_{\mu\nu}^+ \rho_{T\mu}^- - W_{\mu\nu}^- \rho_{T\mu}^+)Z^0_\nu] \nn\\
&&- \frac{i g^2 fD^2}{2\sqrt{2}\,g_T B^2\cos\thw}(\rho_{T\mu\nu}^+ W_\mu^- -
\rho_{T\mu\nu}^- W_\mu^+)Z^0_\nu \nn\\
&& -\frac{egg_T(B+D)\Ntc y_1\tan\thw}{32\sqrt{2} B\,\pi^2}
(\rho_{T\mu}^+ W_\nu^- + \rho_{T\mu}^- W_\nu^+)\widetilde Z_{\mu\nu}^0
\nn\\
&\longra& -i\grpp\sin^2\chi [\rho_{T\mu}^+ \, \Pi_T^- \olra{\partial_\mu}
\Pi_T^0 + \rho_{T\mu}^- \, \Pi_T^0 \olra{\partial_\mu}\Pi_T^+] \nn\\
&& +\frac{ief D^2 \sin\chi}{2\sqrt{2AB}Bg_T F_1 \sin\thw}
(\rho_{T\mu\nu}^+ W_{\mu\nu}^- - \rho_{T\mu\nu}^- W_{\mu\nu}^+)\Pi_T^0 \nn\\
&& +\frac{ief D^2 \sin\chi}{\sqrt{2AB}Bg_T F_1 \sin 2\thw}
(\rho_{T\mu\nu}^+ \Pi_T^- - \rho_{T\mu\nu}^-\Pi_T^+) Z_{\mu\nu}^0 \nn\\
&& -\frac{eg_T(B+D)\Ntc y_1 \tan\thw \sin\chi}{32\sqrt{2AB} \,\pi^2 F_1}
(\rho_{T\mu\nu}^+ \Pi_T^- + \rho_{T\mu\nu}^- \Pi_T^+)\widetilde Z_{\mu\nu}^0
\,.
\eea
Note that there is no WZW term in Eq.~(\ref{eq:rpmWpi}).

We include with these the $\tom \ra W^\pm \tpim$, $W^+W^-$,  $Z^0\tpiz$ and
$Z^0 Z^0$ operators, all of which arise from $\LWZW$:
\bea\label{eq:omWpi}
\CL(\tom \ra W^\pm \tpimp) &\cong&
-\frac{eg_T(B+D)\Ntc \cos\chi}{64\sqrt{2AB} \,\pi^2 F_1 \sin\thw}\, \omega_{T\mu\nu}
(\widetilde W_{\mu\nu}^+ \tpim + \widetilde W_{\mu\nu}^- \tpip) \,; 
\eea
\bea\label{eq:omWW}
\CL(\tom \ra W^+ W^-) &\cong&
\frac{eg_T(B+D)\Ntc}{64\sqrt{2}B \,\pi^2 \sin\thw}\,
\omega_{T\mu} (W_\nu^- \widetilde W_{\mu\nu}^+  + W_\nu^+ \widetilde
W_{\mu\nu}^-) \nn\\
&\longra& 
\frac{eg_T(B+D)\Ntc \sin\chi}{64\sqrt{2AB} \,\pi^2 F_1
  \sin\thw}\,\omega_{T\mu\nu}
(\widetilde W_{\mu\nu}^+ \Pi_T^- + \widetilde W_{\mu\nu}^- \Pi_T^+)
\,; 
\eea
\bea\label{eq:omZpi}
\CL(\tom \ra Z^0 \tpiz) &\cong&
-\frac{e g_T(B+D)\Ntc \cot 2\thw \cos\chi}{32\sqrt{2AB} \,\pi^2 F_1}\,
\omega_{T\mu\nu} \widetilde Z_{\mu\nu}^0 \tpiz\,;
\eea
\bea\label{eq:omZZ}
\CL(\tom \ra Z^0 Z^0) &\cong& 
\frac{g^2g_T(B+D)\Ntc \cos 2\thw}{64\sqrt{2}B \,\pi^2 \cos^2\thw}\, \omega_{T\mu}
Z_\nu^0 \widetilde Z_{\mu\nu}^0\nn\\
&\longra& \frac{e g_T(B+D)\Ntc \cot 2\thw \sin\chi}{32\sqrt{2AB} \,\pi^2 F_1}\,
\omega_{T\mu\nu} \widetilde Z_{\mu\nu}^0 \Pi_T^0 \,.
\eea

Next, we list phenomenologically interesting couplings of $\ta$ to $\tpi$
and a weak boson. To the order we calculated these, the parity-violating
decays $\ta \ra \tpi W_L$ do not occur.
\bea\label{eq:apmZpi}
\CL(\tapm \ra Z^0\tpipm) &\cong& \frac{i gg_T \sqrt{B}C F_1\cos 2\thw \cos\chi}
{4\sqrt{2A}\cos\thw}(a_{T\mu}^+\tpim - a_{T\mu}^-\tpip)Z^0_\mu \nn\\
&&-\frac{ig C\cos 2\thw \cos\chi}{\sqrt{2AB}\, g_T F_1 \cos\thw}
[(a_{T\mu\nu}^+ \partial_\mu\tpim - a_{T\mu\nu}^- \partial_\mu\tpip)Z^0_\nu \nn\\
&& \qquad\qquad\qquad\qquad +(a_{T\mu}^+\partial_\nu\tpim -
a_{T\mu}^-\partial_\nu\tpip)Z^0_{\mu\nu}]
 \nn\\
&&-\frac{ig fD \cos 2\thw \cos\chi}{\sqrt{2AB}g_T F_1 \cos\thw}
(a_{T\mu}^+\partial_\nu\tpim - a_{T\mu}^-\partial_\nu\tpip)Z^0_{\mu\nu} \nn\\
&\cong& -\frac{iefD \cot 2\thw \cos\chi}{\sqrt{2AB} g_T F_1}(a_{T\mu\nu}^+
\tpim - a_{T\mu\nu}^- \tpip)Z_{\mu\nu}^0 \,.
\eea
This decay is of particular interest at the LHC because it is so far the only
one involving a technipion that has been shown to be visible above
backgrounds at the LHC~\cite{LSTCazuelos,Brooijmans:2008se}. Note that, in
our $\Leff$, only the $f$-term has the derivative structure to contribute to
this and other $\ta$ decays of interest. Similarly,
\bea\label{eq:apmWpi}
\CL(\tapm \ra W^\pm \tpiz) &\cong& 
\frac{iefD\cos\chi}{2\sqrt{2AB}g_T F_1\sin\thw} (a_{T\mu\nu}^+ W_{\mu\nu}^- -
a_{T\mu\nu}^- W_{\mu\nu}^+)\tpiz\,;
\eea
and
\bea\label{eq:azWpi}
\CL(\taz \ra W^\pm \tpimp) &\cong& 
\frac{iefD\cos\chi}{2\sqrt{2AB}g_T F_1 \sin\thw}\, a_{T\mu\nu}^0 
(W_{\mu\nu}^+ \tpim - W_{\mu\nu}^- \tpip)\,,
\eea
and
\bea\label{eq:azWW}
\CL(\taz \ra W^+ W^-) &\cong& 
-\frac{ig^2fD}{2\sqrt{2}g_TB}\, a_{T\mu}^0 (W_{\mu\nu}^+ W_\nu^- -
W_{\mu\nu}^- W_\nu^+)\nn\\
&\longra& -\frac{iefD \sin\chi}{2\sqrt{2AB} g_T F_1 \sin\thw}\,
a_{T\mu\nu}^0 (W_{\mu\nu}^+ \Pi_T^- - W_{\mu\nu}^- \Pi_T^+)\,.
\eea
Finally, the decay $\tapm \ra W^\pm Z^0$ may be sought in the $\tropm \ra
W^\pm Z^0$ analysis:
\bea\label{eq:apmWZ}
\CL(\tapm \ra W^\pm Z^0) &\cong& \frac{i g^2 fD}{2\sqrt{2}\,g_T B\cos\thw}
[(a_{T\mu}^+ W_\nu^- - a_{T\mu}^- W_\nu^+)Z^0_{\mu\nu}\cos 2\thw\nn\\
&&\qquad\qquad\qquad\,\,\, + (W_{\mu\nu}^+ a_{T\mu}^-  - W_{\mu\nu}^-
a_{T\mu}^+)Z^0_\nu] \nn\\
&\longra& \frac{iefD\cot 2 \thw\sin\chi} {\sqrt{2AB}g_T F_1}
(a_{T\mu\nu}^+ \Pi_T^- - a_{T\mu\nu}^-  \Pi_T^+)Z^0_{\mu\nu} \nn\\
&& -\frac{iefD\sin\chi}{2\sqrt{2AB}g_T F_1\sin\thw}
(a_{T\mu\nu}^+ W_{\mu\nu}^- - a_{T\mu\nu}^- W_{\mu\nu}^+)\Pi_T^0\,.
\eea

The $\tapm$ radiative decays, which were an important motivation for
including the $f$-interaction in Eq.~(\ref{eq:Lsigone}) are listed next.
\bea\label{eq:apmgampi}
\CL(\tapm \ra \gamma \tpipm) &\cong& -\frac{ie fD\cos\chi}{\sqrt{2AB}\,g_T
  F_1}\, (a_{T\mu\nu}^+ \tpim - a_{T\mu\nu}^- \tpip) F_{\mu\nu} \,; 
\eea
and
\bea\label{eq:apmgamW}
\CL(\tapm \ra \gamma W^\pm) &\cong& -\frac{ieg fD}{\sqrt{2}\,g_T
  B}\,(a_{T\nu}^+ W_\mu^- - a_{T\nu}^- W_\mu^+)F_{\mu\nu} \nn\\
&\longra& \frac{ie fD\sin\chi}{\sqrt{2AB}\,g_T F_1}\, (a_{T\mu\nu}^+ \Pi_T^-
- a_{T\mu\nu}^- \Pi_T^+) F_{\mu\nu} \,.
\eea
Here, $F_{\mu\nu} = \partial_\mu A_\nu - \partial_\nu A_\mu$, where $A_\mu$
is the photon field.

Finally, there are the interactions for the main radiative decays of $\omega$
and $\tro$. They arise from the WZW terms in the effective Lagrangian.
\bea\label{eq:omgampi}
\CL(\tom \ra \gamma \tpiz) &\cong&
-\frac{eg_T\Ntc(B+D)\cos\chi}{32\sqrt{2AB}\,\pi^2 F_1}\,
\omega_{T\mu\nu}\widetilde F_{\mu\nu}\tpiz\,;
\eea
and
\bea\label{eq:omgamZ}
\CL(\tom \ra \gamma Z^0) &\cong& \frac{eg g_T \Ntc (B+D)}
  {32\sqrt{2}\,\pi^2 B\cos\thw}\, \omega_{T\mu}Z^0_\nu \widetilde F_{\mu\nu}
  \nn\\
&\longra& \frac{eg_T \Ntc (B+D)\sin\chi}{32\sqrt{2AB}\,\pi^2 F_1}\,
  \omega_{T\mu\nu} \widetilde F_{\mu\nu}\Pi_T^0\,;
\eea
and
\bea\label{eq:rzgampi}
\CL(\troz \ra \gamma \tpiz) &\cong& -\frac{e g_T \Ntc
(B+D) y_1(1+2\sin^2\thw) \cos\chi}{32\sqrt{2AB}\,\pi^2 F_1}
\, \rho_{T\mu\nu}^0\widetilde F_{\mu\nu}\tpiz\,;
\eea
and
\bea\label{eq:rzgamZ}
\CL(\troz \ra \gamma Z^0) &\cong& \frac{eg g_T \Ntc (B+D) y_1(1+2\sin^2\thw)}
  {32\sqrt{2}\,\pi^2 B\cos\thw}\rho^0_{T\mu}Z^0_\nu \widetilde F_{\mu\nu} \nn\\
&\longra& \frac{eg_T \Ntc (B+D) y_1(1+2\sin^2\thw)
  \sin\chi}{32\sqrt{2AB}\,\pi^2 F_1}\,
  \rho_{T\mu\nu} \widetilde F_{\mu\nu}\Pi_T^0\,.
\eea
For $\tropm$,
\bea\label{eq:rpgampi}
\CL(\tropm \ra \gamma \tpipm)  &\cong& -\frac{e g_T \Ntc (B+D)y_1\cos\chi}
{32\sqrt{2AB}\,\pi^2 F_1}
\, [\rho_{T\mu\nu}^+ \tpim + \rho_{T\mu\nu}^- \tpip]\widetilde F_{\mu\nu}\,;
\eea
and
\bea\label{eq:rpgamW}
\CL(\tropm \ra \gamma W^\pm) &\cong& \frac{eg g_T \Ntc (B+D)y_1}
  {32\sqrt{2}\,\pi^2 B}[\rho^+_{T\mu}W^-_\nu + \rho^-_{T\mu}W^+_\nu]
  \widetilde F_{\mu\nu} \nn\\ 
&\longra& 
\frac{eg_T \Ntc (B+D) y_1 \sin\chi}{32\sqrt{2AB}\,\pi^2 F_1}\,
 [\rho_{T\mu\nu}^+ \Pi_T^- + \rho_{T\mu\nu}^- \Pi_T^+]\widetilde F_{\mu\nu}
\,.
\eea

Let us now compare these decay operators with the TCSM amplitudes in App.~A
(see Eqs.~(\ref{eq:amplitudes},\ref{eq:VA}) and the table of $V_{V_T/\ta
G_\perp \tpi}$ and $A_{V_T/\ta G_\perp \tpi}$ factors). Although not listed in
that table, it is clear that the operators for $\tro \ra \tpi\tpi$,
$W_L\tpi$ and $W_L W_L$ are consistent with both isospin symmetry and the
replacement of $\cos\chi$ by $\sin\chi$ for each replacement of $\tpi$ by
$W_L$. (Actually, there is a peculiar change in the sign of $\sin\chi$ relative
to the TCSM, but this has no observable consequence.)

Consider the WZW interactions above. The mass scale suppressing these
interactions is (using the notation of App.~A)
\be\label{eq:WZWscale} M_{V_1} = \frac{16\sqrt{2AB}\,\pi^2F_1}{g_T\Ntc(B+D)}
\cong \frac{4\pi\sqrt{2AB/b} M_{\tro}}{\atro\Ntc (B+D)}\,.  \ee
Using the estimate in Eq.~(\ref{eq:xsqest}) and assuming all $\Lsig$
strengths are $\CO(1)$, we have $M_{V_1} \simeq 2 M_{\tro}$. This estimate is
in reasonable accord with QCD where the corresponding mass for $\rho,\omega
\ra \gamma \pi^0$ is $M_V \simeq 700\,\mev \simeq M_\rho$. Moreover, it would
not be surprising to find that $\atro$ is twice as large as the naive scaling
from QCD in Eq.~(\ref{eq:xsqest}) suggests, so that $M_{V_1} \simeq M_{\tro}$
then.

The strengths of these WZW interactions agree with the amplitude factors
$V_{V_T G_\perp \tpi}$ in App.~A except for two pairs of operators. In the
TCSM table, $V_{\troz W_\perp^\pm \tpimp} = V_{\troz W_\perp^\pm W_L^\mp} =
0$, but they are proportional to $y_1\sin\thw$ in
Eqs.~(\ref{eq:rzWpi},\ref{eq:rzWW}). And $V_{\troz Z_\perp^0 \tpiz}$,
$V_{\troz Z_\perp^0 Z_L^0} \propto -y_1\tan\thw$ in the TCSM, but they are
proportional to $-2y_1\sin^2\thw\tan\thw$ in
Eqs.~(\ref{eq:rzZpi},\ref{eq:rzZZ}). The reason for these discrepancies is
this: In the TCSM~\cite{Lane:2002sm}, no mixing was allowed between $\troz$
and $\tom$ because, it was argued, this mixing is negligibly small in QCD.
This argument is plausible {\em unless} the zeroth-order $\troz$ and $\tom$
--- our $V_3$ and $V_0$ --- are degenerate. And that is exactly what happens
in our HLS model with equal $SU(2)_{L,R}$ and $U(1)_{L,R}$ couplings; see
$M_0^2$ in Eq.~(\ref{eq:Mzero}). Thus, any amount of nonzero mixing can have
a significant, $\CO(x^0)$, effect. From the eigenvectors in
Eq.~(\ref{eq:neutevecs}) of App.~B, We see that
\be\label{eq:rhoomegamix}
|\troz\rangle \cong \cos\epsilon\, |V_3\rangle + \sin\epsilon\,
|V_0\rangle\,,\qquad
|\tom\rangle \cong -\sin\epsilon\, |V_3\rangle + \cos\epsilon\, |V_0\rangle\,;
\ee
where $\epsilon \cong 2y_1\sin^2\thw$ is presumed small compared to
one.\footnote{This is an instructive problem in degenerate perturbation
  theory. See K.~Gottfried, {\em Quantum Mechanics: Fundamentals}, First
  Edition, 1966, p.~397, Problem~1, and J.~J.~Sakurai, {\em Modern Quantum
    Mechanics}, Revised Edition, 1993, p.~348, Problem~5.12. The exact result
  for the mixing angle is $\epsilon = \half\tan^{-1}(4y_1
  \sin^2\thw/(1-4y_1^2 \sin^2\thw))$.} Thus, $V_{\troz\, W_\perp\, \tpi/W_L}
\cong V_{V_3\, W_\perp\, \tpi/W_L} + 2y_1 \sin^2\thw V_{V_0\, W_\perp\,
\tpi/W_L}$, where the amplitude factors for $V_3$ and $V_0$ are in App.~A.
Since the $V_0$ amplitude factors in the table are of zeroth order in
$y_1\sin\thw$, they are equal, to the order in~$\epsilon$ we calculated, to
the corresponding $\tom$ decay strengths we found in
Eqs.~(\ref{eq:omWpi}--\ref{eq:omZZ}).

All the $\rho_{T\mu\nu}G_{\perp\,\mu\nu}\tpi$ and
$a_{T\mu\nu}G_{\perp\,\mu\nu}\tpi$ operators are also in accord with the
corresponding TCSM $A_{\tro/\ta\ G_\perp \tpi}$ factors (up to that pesky
sign of $\sin\chi$). We deduce that the mass scale $M_{A_1}$ suppressing
$\rho_{T\mu\nu}G_{\perp\mu\nu}\tpi$ operators is
\be\label{eq:FFscaleone}
M_{A_1} \equiv \frac{\sqrt{2AB}Bg_T F_1}{2fD^2}
= \frac{g_T B^2 F_\pi \sin\chi}{\sqrt{2} f D^2}
\cong \frac{\sqrt{2A}BM_{\ta}}{fD^2}\,.
\ee
For $\Lsig$ couplings of $\CO(1)$, $M_{A_1} \simeq M_{\ta}\simeq M_{\tro}$,
which is what we would naively expect. The characteristic scale suppressing
the $\ta$-decay interactions is
\be\label{eq:FFscaletwo}
M_{A_2} \equiv \frac{\sqrt{2AB}g_T F_1}{2fD} 
= \frac{g_T B F_\pi \sin\chi}{\sqrt{2} f D}
\cong \frac{\sqrt{2A}M_{\ta}}{fD}\,.
\ee
For $\Lsig$ couplings of $\CO(1)$, $M_{A_2} \simeq M_{\ta}$. Note, from
Eq.~(\ref{eq:FFscaleone}), that $M_{A_1} = M_{A_2}$ when $B=D$, i.e., $2c + d
= 0$. This is almost the same as the condition $c=d=0$ for $M_{\tro} =
M_{\ta}$. It is quite remarkable, despite our uncertainty about the theory
underlying our $\Leff$, how closely it tracks the
valence-quark-model-inspired TCSM. Naive dimensional analysis fixes the
TCSM's arbitrary mass parameters $M_{V_i}$ and $M_{A_i}$ in
Eqs.~(\ref{eq:WZWscale},\ref{eq:FFscaleone},\ref{eq:FFscaletwo}) to be what
one would expect merely by scaling from QCD.

\section*{VII. $F_1$-Scale Contribution to $S$, $T$, $W$, $Y$}

In this section we present the results of a calculation of the $F_1$-scale
contribution to the precision electroweak parameters $S$, $T$, $W$ and
$Y$~\cite{Peskin:1990zt,Maksymyk:1993zm,Barbieri:2004qk, Marandella:2005wd}.
The complete parameters are defined in terms of the technicolor contribution
to the polarization functions of electroweak currents as follows:
\bea\label{eq:STWYdefs}
&& S = 16\pi (\Pi_{33}'(0) - \Pi_{3Q}'(0)) \,,\quad 
T  = \frac{16\pi}{M_Z^2\sin^2 2\thw}(\Pi_{11}(0) - \Pi_{33}(0))\nn\\
&& W = {\half} g^2 M_W^2 \Pi''_{3 3}(0)\,,\qquad
Y = {\half} g^{\prime 2} M_W^2 \Pi''_{YY}(0)\,,
\eea
where $\Pi'(0) = (d\Pi(q^2)/dq^2)_{q^2=0}$. Barbieri, {\em et al.}, argued
that these four quantities describe the most important effects on
standard-model processes at energies well below the technivector masses. They
are well-constrained by $e^+e^-$ data at the $Z^0$ and
above~\cite{Amsler:2008zzb,Barbieri:2004qk}:
\bea\label{eq:STWYvals}
&& S = -0.04 \pm 0.09\,(-0.07)\,,\qquad 
T =  0.02 \pm 0.09\,( +0.09) \nn\\
&& W = (-2.7 \pm 2.0)\times 10^{-3}\,,\qquad\,
Y = (4.2 \pm 4.9)\times 10^{-3}\,,
\eea
where, for $S$ and $T$, $U=0$ was assumed and the central value corresponds
to subtracting out the contribution of a standard model Higgs boson of mass
$117\,\gev$; the correction to the central value when the Higgs mass is
increased to $300\,\gev$ is given in parentheses.

To calculate $S_1,\dots,Y_1$, we follow the method described in
Refs.~\cite{Barbieri:2004qk, Marandella:2005wd}. It applies here because
quarks and leptons couple only to {\em primordial} ${\bs W}$ and $B$ and only
in the standard way. We use the technivectors' lowest-order equations of
motion to integrate them out of $\Pi(q^2)$ and then canonically re-normalize
the $W^{\pm,0}$ and $B$ fields' kinetic terms by dividing the fields by the
square roots of
\be\label{eq:WBnorms}
\CN_W = 1 + \frac{(B^2 + D^2)x^2}{B^2}\,,\quad
\CN_B = 1 + \frac{(B^2(1+4 y_1^2) + D^2)x^2\tan^2\thw}{B^2} \,,
\ee
where $A, \dots, D$ were defined in Eq.~(\ref{eq:etaGfactors}). We obtain, to
leading order in $M_W^2/M_{\tro}^2$ and $x^2$:
\bea\label{eq:STWYresults}
S_1 &=& \frac{8\pi (B^2-D^2)}{g_T^2 B^2}\\
T_1 &=& 0 \\
W_1 &=& \frac{x^4(B^3 + b D^2)}{2bB^3}\biggl[2\biggl(\frac{F_2^2}{F_1^2} + a +
b\biggr) - b - \frac{D^2}{B}\biggr]\simeq \frac{x^4 F_2^2}{b F_1^2}\biggl[1 +
\frac{bD^2}{B^3}\biggr]\\
Y_1 &=& \frac{x^4(B^3(1+4y_1^2) + b D^2)\tan^2\thw}{2bB^3}
\biggl[2\biggl(\frac{F_2^2}{F_1^2} + a + b\biggr) - b -
  \frac{D^2}{B}\biggr]\nn\\
&\simeq& \frac{x^4 F_2^2\tan^2\thw}{b F_1^2}\biggl[1 + 4y_1^2 +
\frac{bD^2}{B^3}\biggr]\,.
\eea
Leave $S_1$ aside for a moment. That $T_1 = 0$ (and $U_1 = 0$) at tree level
is guaranteed by the model's built-in custodial isospin symmetry. For
$|A|,\dots,|D| = \CO(1)$ and the estimate $x^2 \simeq 0.5 \times 10^{-2}$ we
made in Sec.~III, $W_1$ and $Y_1 = \CO(x^4)$ are well within experimental
bounds.

Regarding $S_1$, first note that it is what we would get from
Eq.~(\ref{eq:Snarrow}) using Eq.~(\ref{eq:grelations}). Then under the same
assumptions on $B$ and $D$, $S_1$ is likely to be an order of magnitude too
large. However, we can make $S_1$ small by choosing $B^2 \cong D^2$.
Positivity of $M^2_{\tro}$ and $M^2_{\ta}$ require $b,B > 0$. Then, it seems
likely that $D = b+d > 0$ also, so that the condition for small $S_1$ is
\be\label{eq:Czero}
C = B - D = 2c + d \cong 0\,.
\ee
This is implied by the condition $c=d=0$ assumed in
Refs.~\cite{Casalbuoni:1995qt,Appelquist:1999dq} to make $S$ small in their
models. In those references, $C = 0$ implies the vanishing of $\grpp$ so that
$W_L W_L$ scattering in the $J=1$ channel has no $\tro$ pole to unitarize it.
For us, the $f$-term in $\Lsig$ gives $\grpp \cong bBf g_T/4\sqrt{2}A$ and a
$\tro$ coupling to $W_L W_L$ of $\grpp\sin^2\chi$.\footnote{Another way to
  make $S_1$ small is to note that, if $\grpp$ is held fixed, a decrease of
  $B^2-D^2$ by a multiplicative factor $\epsilon < 1$ is compensated by $g_T
  \ra g_T/\epsilon$ and $f \ra \epsilon f$. This decreases $S_1$ by
  $\epsilon^3$. It is also possible, of course, that $S_1$ is not small, but
  is canceled by the contributions to $S$ from other technifermion doublets.
  This course seems less natural to us.}  Finally, if $|C| \ll |B+D|$,
Eqs.~(\ref{eq:mcharev},\ref{eq:mneutev}) imply
\be\label{eq:massratio}
\frac{M^2_{\ta}}{M^2_{\tro}} \simeq \frac{b+d}{b}\,,
\ee
so that the condition $M_{\ta} \simeq M_{\tro}$ further implies that $d$
and, hence, $c$ are both small.

\section*{VIII. Future Projects}

Several projects flow immediately from the effective Lagrangian developed in
this paper. We summarize them here.

 \begin{itemize}

 \item[1.] The ALEPH Collaboration at LEP searched for a $\tro$ enhancement
   in $e^+ e^- \ra W_L^+ W_L^-$ and claimed a limit of $M_{\tro} >
   600\,\gev$~\cite{Schael:2004tq}. Eichten and Lane pointed out that the
   ALEPH analysis does not apply to the TCSM because the $\troz \ra W_L^+
   W_L^-$ coupling is proportional to $\sin^2\chi \ll 1$ and, using a
   simplified version of the HLS model discussed here, showed that ALEPH set
   no meaningful limit on LSTC~\cite{Eichten:2007sx}. That analysis will be
   redone with the $\Leff$ developed here.
  
\item[2.] The HLS effective Lagrangian provides a way to test an assumption
  on which the TCSM relies heavily --- the validity of the approximation
  $W_{L\mu}^\pm \cong \partial_\mu \Pi_T^\pm/M_W =
  2\partial_\mu\Pi_T^\pm/(gF_\pi)$ and the dominance of longitudinally
  polarized weak bosons in such processes as $\tro \ra W \tpi$ and $\ta \ra
  \gamma W$ --- and an important consequence of this approximation, the
  angular distributions in resonant production of $WZ$, $\gamma W$ and
  $\gamma Z$~\cite{Eichten:2007sx}. In a future paper, we shall examine these
  processes and study the $f$-term's effect on them at the resonance mass.

\item[3.] Precision measurements of triple gauge boson vertices at LEP and
  the Tevatron~\cite{Amsler:2008zzb} and, hopefully, soon at the LHC may be
  sensitive to the presence of technivector poles and to the non-standard
  triple gauge boson vertices in the $f$-term of $\Lsig$. These studies at
  the LHC can provide complementary information to the direct technivector
  searches. An analysis of these effects seems worthwhile, therefore. This
  will be a generalization of the study in item~1 above.

\item[4.] If low-scale technicolor is discovered at the LHC, a high energy
  linear $e^+ e^-$ collider such as the ILC or CLIC offers an excellent
  possibility to study the resonant contributions to $e^+e^- \ra
  \ell^+\ell^-$, $W^+W^-$ and $\gamma \tpiz/Z^0$. The energy resolution of
  the collider and its detectors could make it possible to resolve the
  $\troz$, $\tom$ and $\taz$ in their $\ellp\ellm$ decay channels. The linear
  collider is also likely to be the best place to analyze the angular
  distributions in these channels and, perhaps, determine the sum of charges,
  $y_1 = Q_U+Q_D$, of the constituent technifermions. It would be interesting
  to study the $s$-dependence of $W^+W^-$ as it passes through the resonance
  region. And, if backgrounds at the Tevatron and LHC prove daunting, the
  linear collider will be the only place to observe $\tom \ra \gamma \tpiz$,
  an important process because it involves a technipion.

 \end{itemize}

\section*{Acknowledgments} This work has been supported in part by the
U.~S.~Department of Energy Grant~No.~DE-FG02-91ER40676~(KL) and
No.~DE-FG02-92ER-40704~(AM). KL's research was also supported by Laboratoire
d'Annecy-le-Vieux de Physique des Particules (LAPP) and Laboratoire
d'Annecy-le-Vieux de Physique Theorique (LAPTH) and he is grateful to LAPTH
for its hospitality. We also thank Barry Barish, Fawzi Boudjema, Estia
Eichten, Chris Hill, Richard Hill and Eric Pilon for valuable discussions.

\vfil\eject

\section*{Appendix~A: Table of TCSM Couplings}

{\begin{table}[!ht]
    \begin{center}{
 \begin{tabular}{|c|c|c|}
 \hline
 Process & $V_{V_T/a_T G_\perp\tpi}$ & $A_{V_T/a_T G_\perp\tpi}$\\
 \hline\hline
 $\tom \ra \gamma \tpiz$& $\cos\chi$ & 0\\
 $\ts\ts\ts\quad \ra \gamma Z^0_L$ & $\sin\chi$ & 0\\ 
 $\ts\ts\ts\qquad \ra W^\pm_\perp \tpimp$ & $\cos\chi/(2\sin\thw)$ & 0 \\ 
  $\,\,\,\;\;\qquad \ra W^\pm_\perp W_L^\mp$ & $\sin\chi/(2\sin\thw)$ & 0 \\
 $\qquad \ra Z^0_\perp \tpiz$ & $\cos\chi\cot 2\thw$ & 0 \\ 
  $\,\qquad \ra Z^0_\perp Z_L^0$ &$\sin\chi\cot 2\thw$ & 0\\
 \hline
 $\troz \ra \gamma \tpiz$ & $y_1\cos\chi$ & 0 \\
 $\; \ra \gamma Z_L^0$ & $y_1\sin\chi$ & 0 \\
 $\;\;\quad \ra W^\pm_\perp \tpimp$ & 0 & $\pm\cos\chi/(2\sin\thw)$\\ 
  $\;\;\;\quad \ra W^\pm_\perp W_L^\mp$ & 0 & $\pm\sin\chi/(2\sin\thw)$\\
 $\;\quad\ra Z^0_\perp \tpiz$ & $-y_1\ts \cos\chi\tan\thw$ & 0\\
  $\;\quad \ra Z^0_\perp Z_L^0$ & $-y_1\sin\chi\tan\thw$ & 0\\
 \hline
 $\tropm \ra \gamma \tpipm$ & $y_1\cos\chi$ & 0\\ 
 $\;\;\;\ra \gamma W_L^\pm$ & $y_1\sin\chi$ & 0\\
 $\;\;\; \ra Z^0_\perp \tpipm$ & $-y_1\cos\chi\tan\thw$ & $\pm \cos\chi
 /(\sin 2\thw)$\\  
  $\quad\;\; \ra Z^0_\perp W_L^\pm$ & $-y_1\sin\chi\tan\thw$ &
  $\pm\sin\chi/(\sin2\thw)$\\
 $\quad\;\; \ra W^\pm_\perp \tpiz$ & 0 & $\mp\cos\chi/(2\sin\thw)$\\ 
  $\quad\;\; \ra W^\pm_\perp Z_L^0$ & 0 & $\mp\sin\chi/(2\sin\thw)$\\
 \hline
  $a_T^0\ra W^\pm_\perp \tpimp$ & 0 & $\mp\cos\chi/(2\sin\thw)$ \\ 
  $\qquad \ra W^\pm_\perp W_L^\mp$    & 0 & $\mp\sin\chi/(2\sin\thw)$\\
  \hline
  $a_T^\pm \ra \gamma \tpipm$ & 0 & $\mp\cos\chi$ \\
  $\qquad \ra \gamma W_L^\pm$ & 0 & $\mp\sin\chi$\\
  $\qquad\;\, \ra W^\pm_\perp \tpiz$ & 0 & $\pm\cos\chi/(2\sin\thw)$\\ 
  $\quad\;\;\; \ra Z_L^0\tpipm$ & 0 & $\mp\cos\chi\cot 2\thw$\\
  $\,\;\qquad \ra W^\pm_\perp Z_L^0$ & 0 & $\pm\sin\chi/(2\sin\thw)$\\ 
  $\,\;\qquad \ra W_L^\pm Z_\perp^0$ & 0 & $\mp\sin\chi\cot 2\thw$ \\ 
  \hline\hline
\end{tabular}}
%
%
\end{center}
\end{table}}

The table above presents the amplitude factors in the TCSM for $V_T (=
\tro,\tom)$ and $\ta$ decay into a technipion plus a transversely-polarized
electroweak boson or one transverse and one longitudinal electroweak
boson~\cite{Lane:2002sm,Eichten:2007sx}. The amplitudes are defined in terms
of the following matrix elements:
\be\label{eq:amplitudes}
\CM(V_T/a_T(p_1) \ra G(p_2) \tpi(p_3)) = \frac{eV_{V_T,a_T\, G_A
    \pi_T}}{2M_{V_{1,2}}}
\,\widetilde F_1^{\lambda\mu}F^*_{2\lambda\mu} + \frac{eA_{V_T,a_T\, G_V
    \pi_T}}{2M_{A_{1,2}}}
\,F_1^{\lambda\mu}F^*_{2\lambda\mu}\,.\\
%
%
\ee
Here, $F_{n\lambda\mu} = \epsilon_{n\lambda} \, p_{n\mu} - \epsilon_{n\mu} \,
p_{n\lambda}$ and $\widetilde F_{n\lambda\mu} = \half
\epsilon_{\lambda\mu\nu\rho} F_n^{\nu\rho}$. The TCSM mass parameters
$M_{V_1}$ and $M_{A_1}$ are expected to be of order $M_{\tro} \cong M_{\tom}$
while and $M_{V_2},M_{A_2} = \CO(M_{\ta})$. The factors $V_{V_T,a_T\, G
  \pi_T}$ and $A_{V_T,a_T\, G_V \pi_T}$ are given by:
\bea\label{eq:VA}
V_{V_T G_\perp\tpi} &=& 2\,{\rm Tr}(Q_{V_T} \{Q^\dagger_{G_V}, \,
Q^\dagger_{\tpi}\}) \,,\qquad
A_{V_T G_\perp\tpi} = 2\,{\rm Tr}(Q_{V_T} [Q^\dagger_{G_A}, \,
Q^\dagger_{\tpi}]) \,;\\
V_{a_T G_\perp\tpi} &=& 2\,{\rm Tr}(Q_{a_T} \{Q^\dagger_{G_A}, \,
Q^\dagger_{\tpi}\}) \,,\qquad
A_{a_T G_\perp\tpi} = 2\,{\rm Tr}(Q_{a_T} [Q^\dagger_{G_V}, \,
Q^\dagger_{\tpi}]) \,.
\eea
In the TCSM, with electric charges $Q_U$, $Q_D$ for $T_U$, $T_D$, and $y_1 =
Q_U + Q_D$, the generators $Q$ in Eq.~(\ref{eq:VA}) are given by
\bea\label{eq:charges}
Q_{\tom} &=& \left(\ba{cc} \half & 0 \\ 0 & \half \ea\right)\nn\\
Q_{\troz,\taz} &=& = \left(\ba{cc} \half & 0 \\ 0 & -\half \ea\right)
\ts;\qquad
Q_{\trop,\tap} = Q^\dagger_{\trom,\tam} =
{1\over{\sqrt{2}}}\left(\ba{cc} 0 & 1 \\ 0 & 0 \ea\right)\nn\\
Q_{\tpiz} &=& \cos\chi \left(\ba{cc} \half & 0 \\ 0 & -\half \ea\right)
\ts;\ts\quad
Q_{\tpip} = Q^\dagger_{\tpim} = {\cos\chi\over{\sqrt{2}}}
 \left(\ba{cc} 0 & 1 \\ 0 & 0  \ea\right)\nn\\
Q_{\tpipr} &=& \cos\chipr \left(\ba{cc} \half & 0 \\ 0
  & \half \ea\right) \nn\\
Q_{\gamma_V} &=& \left(\ba{cc} Q_U& 0 \\ 0 & Q_D \ea\right)
\ts;\qquad
Q_{\gamma_A} = 0 \nn\\
Q_{Z_V} &=& {1\over{\sin\thw \cos\thw}} \left(\ba{cc} \fourth - Q_U
  \sin^2\thw & 0 \\ 0 & -\fourth - Q_D \sin^2\thw \ea\right) \nn\\
Q_{Z_A} &=& {1\over{\sin\thw \cos\thw}} \left(\ba{cc} -\fourth & 0 \\ 0
  &  \fourth \ea\right) \nn\\
Q_{W^+_V} &=& Q^\dagger_{W^-_V} = -Q_{W^+_A} = -Q^\dagger_{W^-_A} = {1\over
  {2\sqrt{2}\sin\thw}}\left(\ba{cc} 0 & 1 \\ 0 & 0 \ea\right) \ts.
\eea
%


\section*{Appendix~B: Gauge Boson Mass Eigenstates and Shift Parameters}

The components of the mass eigenstate vectors in the charged sector are
(assuming $c+d$ is not small):
\bea\label{eq:chgevecs}
\hat v_{W^\pm} &=& \biggl\{1 -{\half}\biggl[1 + \frac{D^2}{B^2}\biggr]x^2,\,
x,\, -\frac{Dx}{B}\biggr\}\,,\nn\\
\hat v_{\tropm} &=& \biggl\{-x,\, 1 -\half x^2,\,
\frac{Dx^2}{2(c+d)}\biggr\}\,,\nn\\
\hat v_{\tapm} &=& \biggl\{\frac{Dx}{B},\, -\frac{bDx^2}{2(c+d)B},\, 1 -
\frac{D^2 x^2}{2B^2}\biggr\}\,,
\eea
where the elements are labeled by the primordial gauge bosons $\widehat W^\pm
\equiv (W^1 \mp i W^2)/\sqrt{2}$, $V^\pm$, $A^\pm$. In the same
approximation, the neutral sector eigenvectors, labeled by $W^3$, $B$, $V^3$,
$V^0$, $A_3$, are:\footnote{The exact form of the massless photon eigenvector
  of $M_0^2$ is $\hat v_\gamma = (g'\kappa, g\kappa,
  2xg'\kappa,0,2xy_1g'\kappa)$ where $\kappa = \{(g^2 + g^{'2})[1 +
  4x^2(1+y_1^2)\sin^2\thw]\}^{-1/2}$. This form guarantees that the EM
  current of the standard-model fermions is proportional to $j_{L\mu}^3 +
  j_\mu^Y$.}
%
\bea\label{eq:neutevecs}
\hat v_{\gamma} &=& \biggl\{\sin\thw\biggl[1 -
2x^2(1+y_1^2)\sin^2\thw\biggr],\,
\cos\thw\biggl[1 -
2x^2(1+y_1^2)\sin^2\thw\biggr],\,2x\sin\thw,\,2xy_1\sin\thw,\,0
\biggr\}\,;\nn\\\nn\\
\hat v_{Z} &=& \biggl\{\cos\thw - \frac{x^2}{2\cos\thw}\biggl[1 + \frac{2
  D^2}{B^2} - 4(1 - y_1^2) \sin^4\thw\biggr],\nn\\ 
&\,&\,\,  -\sin\thw\biggl[1+\frac{x^2}{2\cos^2\thw}\biggl(\cos 2\thw
(1+2\cos^2\thw) - \frac{D^2}{B^2} - 4y_1^2\sin^2\thw(1 + \cos^2\thw)\biggr)\biggr],\nn\\
&\,&\,\,  \frac{x\cos 2\thw}{\cos\thw},\, -2xy_1\sin\thw \tan\thw,\,
-\frac{Dx}{B\cos\thw}  \biggr\}\,;\nn\\\nn\\
\hat v_{\troz} &=& \biggl\{-x(1-2y_1^2\sin^4\thw),\, -x\tan\thw \bigl[1 +
2y_1^2\sin^2\thw(1 + \cos^2\thw)\bigr],\nn\\
&\,&\,\, 1 - 2y_1^2\sin^4\thw - \frac{x^2}{2\cos^2\thw}\biggl(1 - 2y_1^2
\sin^4\thw (1 + 4\cos 2\thw)\biggr),\nn\\
&\,&\,\, 2(1-2x^2)y_1\sin^2\thw,\,
-\frac{Dx^2}{2(c+d)\cos^2\thw}\biggl(\cos 2\thw - 2(1 +
2\cos^2\thw)y_1^2 \sin^4\thw\biggr)\biggr\}\,; \nn\\\nn\\
\hat v_{\tom} &=& \biggl\{2xy_1\sin^2\thw,\, -xy_1\sin 2\thw,\,
-2y_1\sin^2\thw\biggl(1 - \frac{x^2(1 + 2\cos 2\thw)}{2\cos^2\thw} \biggr), \nn\\ 
&\,&\,\,  1 - 2y_1^2(\sin^4\thw + x^2\tan^2\thw\cos^2 2\thw),\,
-\frac{2Dx^2 y_1 \sin^2\thw}{c+d}\biggr\}\,;\nn\\\nn\\
\hat v_{\taz} &=& \biggl\{\frac{Dx}{B},\,-\frac{Dx\tan\thw}{B},\,-\frac{bDx^2\cos
  2\thw}{2(c+d)B\cos^2\thw},\, \frac{bD x^2 y_1\tan^2\thw}{(c+d)B},\,
1 - \frac{D^2 x^2}{2 B^2\cos^2\thw}\biggr\}\,.
\eea
Note that all ``mixing angles'' are of $\CO(x^2)$, as would be expected from
the mass matrices in Eqs.~(\ref{eq:Mpm},\ref{eq:Mzero}), {\em except} for
$\troz$--$\tom$. The reason is that their zeroth-order masses are equal so
that the diagonalization of these two states is a problem in degenerate
perturbation theory. The mixing between these two states vanishes entirely
when $y_1 = 0$.

The shift fields $\zeta$ defined in Eq.~(\ref{eq:Gshift}) are given in terms
of the non-canonically normalized $\tilde \pi_T$ by
\bea\label{eq:etaG}
\zeta_{W^\pm} &=& \frac{2AF_1^2\,\tilde\pi^\pm}{(BF_2^2+AF_1^2)} =
2\sin^2\chi\,\tilde\pi^\pm\,,\nn\\
\zeta_{V^\pm} &=& \sqrt{2}\sin^2\chi\,\tilde\pi^\pm\,,\,\,
\zeta_{A^\pm} = -\sqrt{2}\left(\sin^2\chi +
\frac{C}{B}\cos^2\chi\right)\,\tilde\pi^\pm\,; \nn\\
\zeta_{W^3} &=& 0\,,\,\,
\zeta_{B}  = -2\sin^2\chi\,\tilde\pi_3\,,\,\,
\zeta_{V^3} = -\sqrt{2}\sin^2\chi\,\tilde\pi_3\,,\,\,
\zeta_{V^0} = -2\sqrt{2}y_1\sin^2\chi\,\tilde\pi_3\,, \nn\\
\zeta_{A^3} &=& -\sqrt{2}\left(\sin^2\chi +
  \frac{C}{B}\cos^2\chi\right)\,\tilde\pi_3 \,,\,\,
\zeta_{A^0} = -\frac{C\,\tilde\pi_0}{\sqrt{2}B\cos\chi} \,.
\eea

\section*{Appendix~C: Adjustable Parameters in $\Leff$ with \hfill\break
  Suggested Defaults}

We present two schemes for choices of the adjustable parameters in $\Leff$.
The first is the more general and makes essentially no approximations. The
second drops terms of $\CO(x^2)$ in the technivector masses. In all cases,
$F_\pi = \sqrt{F_2^2 + A/B F_1^2} = 2^{-1/4}G_F^{-1/2} \cong 246\,\gev$ is
fixed.

\subsection*{C.1 General Scheme for Parameters with $c+d \neq 0$}

While it would be convenient to use the technivector masses as inputs, this
is is not practical if one wishes to keep the $\CO(x^2)$ terms in their
masses and assume that the $\Lsig$ couplings $c,d$ are not very small; see
Eqs.~(\ref{eq:mcharev},\ref{eq:mneutev}). In this case, we recommend the
following choice of independent input parameters:
\bea\label{eq:paramsetone}
&& a,b,c,d,f; \nn\\
&& g_T,y_1,\Ntc,\sin\chi = \sqrt{A/B}F_1/F_\pi; \nn\\
&& {\rm {quark\,\,and\,\,lepton\,\,masses\,\,(at\,\,}}M_{\tpi}): m_{u_i},
m_{d_i}, m_{\ell_i}, \,\, i=1,2,3\,; \nn\\
&& {\rm technipion\,\,masses\,\,}M_{\tpi} \equiv M_{\tpipm} = M_{\tpiz} =
M_1,\,\, M_{\tpipr} = \sqrt{M_1^2 + M_2^2}\,; \nn\\
&&\tpi\,\,{\rm couplings\,\,to\,\,quarks\,\,and\,\,leptons;
  \,\,see\,\,Eq.~(\ref{eq:Lpiff})}\,.  \eea
In terms of these, $A = aB + bc -\thalf d^2$, $B = b + 2(c+d)$, $C= 2c+d$, $D
= b+d = B - C$, $x^2 = g^2/2g_T^2$, $F_1 = \sqrt{B/A} F_\pi \sin\chi$, and
$\grpp = b(B^2 + (f-1)D^2)g_T/(4\sqrt{2}AB)$.

In general, one would experiment with these parameters to determine a set
that gives the desired $\tro$, $\tom$ and $\ta$ masses. Suggested defaults,
corresponding roughly to recently studied masses using {\sc Pythia} (and
assuming $a,b,d > 0$), are
\bea\label{eq:defaultsone}
&& a = b = f = 1,\,\, d \simeq \pm 2c = +0.10; \nn\\
&& g_T = \sqrt{8\pi(2.16)(3/\Ntc)}\,\, {\rm with}\,\, \Ntc = 4; \nn\\
&& y_1 = 1\,\, {\rm or}\,\, 0; \nn\\
&&\sin\chi = 1/3 \,\, ({\rm reasonably\,\,}1/4 - 1/2) \nn\\
&& {\rm {quark\,\,and\,\,lepton\,\,masses\,\,at\,\,}}M_{\tpi} {\rm \,\,as\,\,
  in\,\, {\sc Pythia}} \nn\\
&& M_{\tpi} \equiv M_{\tpipm} = M_{\tpiz} = (1/2 - 2/3)\,M_{\troz},\,\,
M_{\tpipr} \simge M_{\taz}\,; \nn\\
&&\tpi\,\,{\rm couplings\,\,to\,\,quarks\,\,and\,\,leptons;
  \,\,see\,\,Eq.~(\ref{eq:Lpiff})}\,.
\eea
These correspond to $M_{\tro} \cong M_{\tom} \cong 258\,\gev$, $M_{\ta} \cong
294\,\gev$, $F_1 = 84\,\gev$, $\grpp = 1.09$, $M_{V_1} = 390\,\gev$, $M_{A_1}
= 517\,\gev$, $M_{A_2} = 437\,\gev$ and $S_1 = 0.18$ for $d = 2c = 0.10$, and
to $M_{\tro} \cong M_{\tom} \cong 269\,\gev$, $M_{\ta} \cong 282\,\gev$, $F_1
= 84\,\gev$, $\grpp = 1.19$, $M_{V_1} = 360\,\gev$, $M_{A_1} = M_{A_2} =
370\,\gev$ and $S_1 = 0$ for $d = -2c = 0.10$; in both cases, $\sin\chi =
\third$ and $y_1 = 1$.

\subsection*{C.2 Scheme for Parameters with $x^2 = 0$ Masses and $c+d \neq 0$}

Since the technivector masses hardly depend on $x^2$ for ``reasonable''
values of $g_T$, we can use them as inputs and solve for some $\Lsig$
parameters in terms of them. We recommend the following choice of independent
input parameters:
\bea\label{eq:paramsettwo}
&& M_{\tro} = M_{\tom} = \thalf g_T F_1 \sqrt{b},\,\, M_{\ta} = \thalf g_T
F_1 \sqrt{B}; \nn\\
&& a,b,f; \nn\\
&& g_T,y_1,\Ntc,\sin\chi = \sqrt{A/B}F_1/F_\pi; \nn\\
&& {\rm {quark\,\,and\,\,lepton\,\,masses\,\,(at\,\,}}M_{\tpi}): m_{u_i},
m_{d_i}, m_{\ell_i}, \,\, i=1,2,3\,; \nn\\
&& {\rm technipion\,\,masses\,\,}M_{\tpi} = M_1,\,\, M_{\tpipr} = \sqrt{M_1^2
  + M_2^2}\,; \nn\\
&&\tpi\,\,{\rm couplings\,\,to\,\,quarks\,\,and\,\,leptons;
  \,\,see\,\,Eq.~(\ref{eq:Lpiff})}\,.
\eea
Solving for the other parameters, we obtain:
\bea\label{eq:otherparamstwo}
&& F_1 = \frac{2M_{\tro}}{g_T \sqrt{b}} = \sqrt{\frac{B}{A}} F_\pi\sin\chi;  \nn\\
&& B = \frac{M_{\ta}^2 b}{M_{\tro}^2} \Lra A \equiv aB + bc - \thalf d^2 = 
\left(\frac{bg_T M_{\ta} F_\pi\sin\chi}{2 M_{\tro}^2}\right)^2;  \nn\\
&& c+d = \thalf(B-b) = \thalf b \left[(\frac{M_{\ta}}{M_{\tro}})^2 -1\right];
\nn\\
&& bc - \thalf d^2 = A-aB \Lra d = -b \pm\sqrt{bB - 2(A-aB)}\,.
\eea
To resolve the quadratic ambiguity, we always take $b > 0$ so that
$M_{\tro}^2 > 0$; then, so long as $b \ge |d|$, so that $D > 0$, the positive
square root is the correct solution. Once $d$ is determined in this way, $c =
\thalf(B+b) - \sqrt{bB - 2(A-aB)}$.

A set of parameters, based on Case~A in Ref.~\cite{Brooijmans:2008se}, is the
following:
\bea\label{eq:defaultstwo}
&& M_{\tro} = M_{\tom} = 300\,\gev, \,\, M_{\ta} = 330\,\gev; \nn\\
&& a = b = f = 1; \nn\\
&& g_T = \sqrt{8\pi(2.16)(3/\Ntc)}\,\, {\rm with}\,\, \Ntc = 4; \nn\\
&& y_1 = 1\,\, {\rm or}\,\, 0; \nn\\
&&\sin\chi = 1/3 \,\, ({\rm reasonably\,\,}1/4 - 1/2) \nn\\
&& {\rm {quark\,\,and\,\,lepton\,\,masses\,\,at\,\,}}M_{\tpi} {\rm \,\,as\,\,
  in\,\, {\sc Pythia}} \nn\\
&& M_{\tpi} \equiv M_{\tpipm} = M_{\tpiz} = (1/2 - 2/3)\,M_{\troz},\,\,
M_{\tpipr} \simge M_{\taz}\,; \nn\\
&&\tpi\,\,{\rm couplings\,\,to\,\,quarks\,\,and\,\,leptons;
  \,\,see\,\,Eq.~(\ref{eq:Lpiff})}\,.
\eea
Taking $\sin\chi = \third$, these lead to $F_1 = 94\,\gev$, $B = 1.21$, $A =
0.92$, $d = 0.34$, $c = -0.23$, $D = 1.34$, $C = -0.13$, $\grpp = 1.48$,
$M_{V_1} = 340\,\gev$, $M_{A_1} = 303\,\gev$, $M_{A_2} = 335\,\gev$ and $S_1
= -0.14$.

\vfil\eject

\bibliography{LSTC_Leff}
\bibliographystyle{utcaps}
\end{document}